# Spinel-bearing Spherules Condensed from the Chicxulub Impact-vapor Plume

Denton S. Ebel*

Department of Earth and Planetary Sciences, American Museum of Natural History, 79th Street at Central Park West, New York, New York 10024-5192, USA

Lawrence Grossman

Department of the Geophysical Sciences and Enrico Fermi Institute, University of Chicago, Chicago, Illinois 60637, USA

**ABSTRACT**

Formation of the giant Chicxulub crater off Mexico's Yucatan Peninsula coincided with deposition of the global Ir-rich Cretaceous-Tertiary (K-T) stratigraphic boundary layer at ca. 65 Ma. The boundary is marked most sharply by abundant spherules containing unaltered grains of magnesioferrite spinel. Here we predict for the first time the sequential condensation of solids and liquids from the plume of vaporized rock expected from oblique K-T impacts. We predict highly oxidizing plumes that condense silicate liquid droplets bearing spinel grains whose compositions closely match those marking the actual boundary. Systematic global variations in spinel composition are consistent with higher condensation temperatures for spinels found at Atlantic and European sites than for those in the Pacific.

**INTRODUCTION**

The iridium spike at the Cretaceous-Paleogene (K-P, formerly K-T) boundary (Alvarez et al., 1980) is vertically diffuse in the sediment column (Kyte and Bostwick, 1995), but the boundary is better defined by a thin, global, spinel-bearing "fireball layer" (Montanari et al., 1983; Smit and Kyte, 1984; Bohor and Glass, 1995; Smit et al., 1992). This consists mostly of abundant spherules <0.5 mm in diameter, containing up to 60 ppb

Ir and unaltered, mostly 1-2 µm, euhedral and dendritic, Ni-bearing magnesioferrite spinel crystals (Montanari et al., 1983; Kyte and Bostwick, 1995). This spherule layer corresponds to the peak of Ir concentration wherever Ir has been measured with high stratigraphic resolution (Kyte and Bostwick, 1995). The origin of the spherules is controversial. They have been interpretted as ablation products from one or more incoming meteors (Gayraud et al., 1996), impact melt droplets (Montanari et al., 1983), or liquid condensates from the Chicxulub impact plume (Kyte and Smit, 1986). It has been argued that an impact-vapor plume would not maintain the relatively oxidizing conditions required to crystallize Fe-rich spinel (Montanari et al., 1983). Here we provide rigorous quantitative evidence to the contrary.

**BOUNDARY-LAYER SPHERULES**

Chemical compositions have been measured for many hundreds of K-P spinels (Table DR1[1]; Kyte and Bostwick, 1995; Kyte and Smit, 1986; Kyte and Bohor, 1995). Individual crystals have Cr decreasing, and $Fe^{3+}/Fe^{2+}$ increasing outward, but the details of intracrystalline zonation are unknown (Preisinger et al., 1997). Spinel composition variability among spherules at each locality is greater than variability within individual spherules (Kyte and Bohor, 1995; Smit and Kyte, 1984). Resistance of magnesioferrite spinels to weathering is suggested by the preservation of primary Ni-rich spinels in Archean spherules (Byerly and Lowe, 1994), the fresh outlines of K-P spinels (Smit and Kyte, 1984), and the lack of evidence for diffusion gradients of Ni within spinels or in adjacent K-P boundary clays (Kyte and Bostwick, 1995).

Spherules condensed from impact fireballs should survive reentry unmelted because they would have much lower intrinsic velocities than high-velocity micrometeorites, particularly for shallow reentry angles (Love and Brownlee, 1991). The lack of depletion of the volatile element Pd in K-P spherules is strong evidence against their melting and partial volatilization (Greshake et al., 1998). Spherules preserve their primary quench textures in many localities. The iron hydroxide + clays observed at all localities record postdepositional in situ partial alteration of all but the spinel fraction of once-liquid droplets.

**METHODS**
**Model Impact-Plume Compositions**

Hydrodynamic models yield proportions of rock types vaporized in oblique and vertical impacts. The hottest part of the plume forms when the front part of the projectile is shock-vaporized on impact and blasted back out through the incoming rear part, which is in turn vaporized. Vaporized rock substantially decouples from less energetic solid and molten rock early in the event, and the entire fireball vapor plume is turbulent and chemically heterogeneous at scales not resolved by the models. The fireball vapor orbits the planet with return trajectories dominated by the rotation and shape of the planet (Argyle, 1989; Kring and Durda, 2002).

Fireball vapor-plume compositions used here (Table DR2 [see footnote 1]) are estimated from hydrodynamic model predictions (Pierazzo and Melosh, 1999, 2000) of target and projectile volumes vaporized in the first 5 s of impact by a 10-km-diameter impactor (50% porosity, 20 km/s velocity) on a 2-km-thick section of shallow shelf sediments overlying continental crust. A CV chondrite composition (Wasson and Kallemeyn, 1988) is assumed for the projectile (Kyte, 1998; Shukolyukov and Lugmair, 1998), and a typical granite is assumed for the crust (G2 of Flanagan, 1967). The impact models consider two different carbonate:anhydrite volumetric ratios in the target rocks, 70:30 and 50:50 (designated C and S in Table 1). The impact models were run for these two target-rock assumptions, at 90° (vertical), 60°, 45°, and 30° incident angles. High-angle (>60°) impacts produce vapors richest in impactor material, the source of Ir. More oblique impacts yield higher proportions of sediment-derived vapor, relative to vapor from crust and impactor. Impacts at <30° yield sediment-rich plumes (Table 1: 30C, 30S) inconsistent with the observed Ir enrichment of the global ejecta layer.

The impact models (Pierazzo and Melosh, 1999, 2000) predict relative volumes of only five vapor components (Table 1, columns 3–7), where "carbonate" combines limestone and dolomite. We averaged data from two published stratigraphic sections (Ward et al., 1995; Sharpton et al., 1996) of Pemex well Y4 near the crater to obtain a thickness of 1928 m and volume fractions of 0.568 limestone (assumed to be pure $CaCO_3$), 0.270 dolomite ($CaMg(CO_3)_2$), 0.112 anhydrite ($CaSO_4$), 0.036 sandstone ($SiO_2$), and 0.013 shale (kaolinite). We used these proportions to modify the vapor compositions predicted by the impact models (Table 1) to account for limestone, dolomite, sandstone, and shale by assuming that the carbonate part has a limestone:dolomite ratio of ~2.1 and by diluting the carbonate- and anhydrite-bearing sediment with sandstone and shale in the well Y4 proportions. Because sulfate and carbonate respond differently to extreme shock, the carbonate:anhydrite ratios from impact models were used instead of the well Y4 ratio. Cases C and S (Table 1) span possible local variations in proportions of thickly bedded sediments not captured by impact models, which could cause plume heterogeneity (Pierazzo and Melosh, 1999, 2000). In separate runs, up to an additional 20 atomic percent of air was mixed with plume compositions to test the effects of possible air entrainment.

**TABLE 1.** VOLUME PROPORTIONS OF VAPOR COMPONENTS.

| Label | Angle | Impactor | Crust | Water | Carbonate | Anhydrite |
|---|---|---|---|---|---|---|
| C: carbonate-rich target | | | | | | |
| 90C | 90° | 0.1418 | 0.1703 | 0.2105 | 0.4116 | 0.0658 |
| 60C | 60° | 0.1182 | 0.1373 | 0.1707 | 0.4942 | 0.0796 |
| 45C | 45° | 0.0461 | 0.0795 | 0.2111 | 0.5655 | 0.0978 |
| 30C | 30° | 0.0194 | 0.0314 | 0.1854 | 0.6525 | 0.1114 |
| S: sulfate-rich target | | | | | | |
| 90S | 90° | 0.1637 | 0.1966 | 0.1737 | 0.3392 | 0.1267 |
| 60S | 60° | 0.1369 | 0.1590 | 0.1411 | 0.4091 | 0.1539 |
| 45S | 45° | 0.0547 | 0.0942 | 0.1789 | 0.4790 | 0.1932 |
| 30S | 30° | 0.0232 | 0.0376 | 0.1587 | 0.5585 | 0.2221 |

**Thermodynamic Calculations**

We modified a chemical thermodynamic code (Ebel and Grossman, 2000; Ebel et al., 2000) to calculate solid + silicate liquid + vapor equilibria along fixed pressure-temperature ($P$-$T$) paths, for various bulk compositions in the system H, C, N, O, Na, Mg, Al, Si, P, S, K, Ca, Ti, Cr, Mn, Fe, Co, and Ni. The gas phase is treated as an ideal equilibrium mixture of 177 ideal gaseous species. The program includes all the nonideal solid solutions and pure solids of Ebel and Grossman (2000). The solution model for spinel $(Mg,Fe^{2+},Al,Fe^{3+},Cr,Ti)_3O_4$ (Sack and Ghiorso, 1991), does not directly address the Ni content (mean ~3 wt%) of the spinels observed at the K-P boundary (Kyte and Bostwick, 1995). Spinel Ni content can, however, be inferred by combining our results with experimental melt-spinel partitioning data. Three model liquids were allowed to condense simultaneously: the MELTS silicate liquid (Ghiorso and Sack, 1995), CaO-MgO-$Al_2O_3$-$SiO_2$ liquid (Berman, 1983), and liquid FeO (Chase, 1995). This liquid approach is a necessity dictated by stoichiometric limitations of existing silicate liquid models and the uniquely Ca-rich impact vapors, but there is good reason to think that the sum of these liquids is not far from the composition of the single liquid that would form. Of course, the Gibbs energies of the end-member liquid components, rather than their mixing properties, exert first-order control on the liquids' thermodynamic stabilities relative to vapor and solid phases.

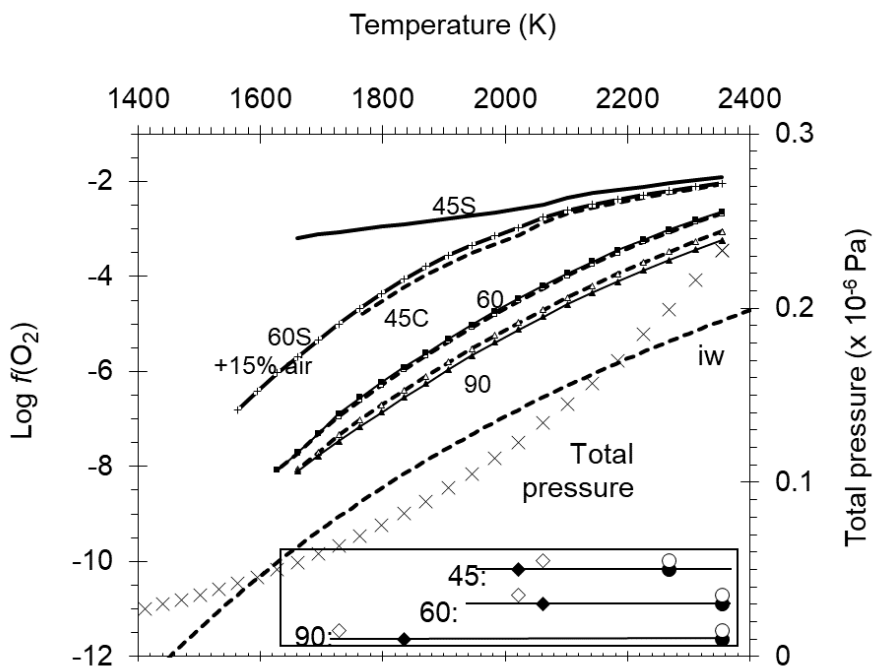

**Figure 1.** Oxygen fugacity [$f(O_2)$; left axis] in cooling vapor plumes for three obliquities and two sediment compositions (solid = S, dashed = C, Table 1) and run 60S with 15% admixed air (+ signs). Curves C and S for 90° and 60° impacts are nearly coincident. Inset lines with symbols record crystallization temperatures of solid spinel (circles) and calcium silicate (diamonds) during cooling of vapor for carbonate-rich cases (filled symbols) and sulfate-rich cases (open symbols). Silicate liquid is always present at 2400 K. Dotted line (iw) shows iron-wüstite (Fe-FeO) oxygen-fugacity buffer curve; curve of "x" symbols shows total pressure (right axis) at each temperature step of calculation.

A rigorous model for the time-pressure-temperature paths followed by heterogeneous parcels of the fireball does not exist. The $P$-$T$ path used here for the cooling vapor (Fig. 1) is for the computed expansion of a sphere of vaporized dunite with an initial internal energy of 71 MJ/kg expanding from a density of 3.2 g·cm$^{-3}$ into vacuum (J. Melosh, 1998, personal commun.). Owing to limits on equations of state, calculations were restricted to the tail end of this cooling path, where $P < 10^6$ Pa and $T < 2400$ K. The plume is expected to cool through this range on time scales of minutes. In every case, depletion of the vapor in refractory elements causes runs to halt close to the temperature of complete liquid crystallization.

**RESULTS AND DISCUSSION**

At high temperature, cooling impact-vapor plumes condense silicate liquid from which Fe-rich spinel, metal alloy, and finally calcium silicates (larnite, $Ca_2SiO_4$; hatrurite, $Ca_3SiO_5$) crystallize sequentially (Fig. 1, inset). Metal alloy always has Ni > 80 wt% because of the highly oxidized nature of the plumes. The predicted change in composition of condensed spinels as temperature decreases is compared with the global spinel database in Figure 2. Spinel compositions are complex, nonlinear functions of the initial vapor composition, and oxygen fugacity, $f(O_2)$, controls the $Fe^{3+}/Fe^{2+}$ ratio in spinel. Spinels from the K-P boundary in the Pacific Ocean are Mg and Al rich, average $Fe^{3+}/Fe^{total} = 0.97 \pm 0.01$, and contain $Fe^{total} = 1.53 \pm 0.12$ and Ni = $0.05 \pm 0.01$ per formula unit (pfu). Atlantic and European K-P spinels average $Fe^{3+}/Fe^{total} = 0.83 \pm 0.03$ and contain $Fe^{total} = 2.02 \pm 0.25$ and Ni = $0.12 \pm 0.03$. The fact that the plumes have $f(O_2)$ well above the Fe-FeO (iron-wüstite [iw]) buffer curve (Fig. 1) causes high $Fe^{3+}/Fe^{total}$ in spinel, even without admixed air. Ni and PGEs (platinum-group elements) would concentrate in condensate spinel grains (Righter and Downs, 2001), particularly at high temperature (Toppani and Libourel, 2003) above the nickel–nickel oxide (NNO) buffer curve, which is nearly coincident with run 45C between 2100 and 1600 K in Figure 1. A spinel/liquid partition coefficient (Wearing, 1983), $[NiO]^{spn}/[NiO]^{liq}$, of five yields 0.06 Ni pfu in spinel for run 60S + 15% air at 1763 K. Spinels can incorporate significant Ir at high $f(O_2)$ (Righter and Downs, 2001). In the magnetic spinel fraction of K-P spherules, which makes up ~0.2 wt% of the bulk sediments (Smit and Kyte, 1984), spinel is a carrier of Ir (Montanari et al., 1983; Bohor and Glass, 1995; Gayraud et al., 1996; Preisinger et al., 1997), and Pt/Ir, Au/Ir, and Pd/Ir ratios are within a factor of two of chondritic. The measured Os/Ir ratio is subchondritic (~0.21 times chondritic), consistent with spinel crystallization near or at equilibrium with the very oxidized, metal-rich vapor we predict, because Os forms a gaseous oxide species more readily than Ir.

No single condensation trajectory can pass through all the spinel data (Fig. 2), so some spatial and temporal variations of the vapor-plume chemistry are required. For sediment-dominated impact vapors, increasing the sulfate:carbonate ratio increases $f(O_2)$ (Fig. 1), because anhydrite contains more oxygen, per mole of vaporized rock, than carbonate. Higher impact angles decrease $f(O_2)$, decrease Mg/Al, and increase Fe/Al in the plume (Table DR2 [see footnote 1]). Without better impact models, it is not possible to uniquely constrain these parameters on the sole basis of the correspondence between our predictions and observed spinel chemistry. Our results do suggest, however, that the spinels found in Atlantic sites record a higher-temperature phase of condensation than those found in the Pacific Ocean. Kyte and Bostwick (1995) concluded the inverse, based on an implied impactor trajectory and a lost high-temperature phase inferred from porous Pacific spinel textures.

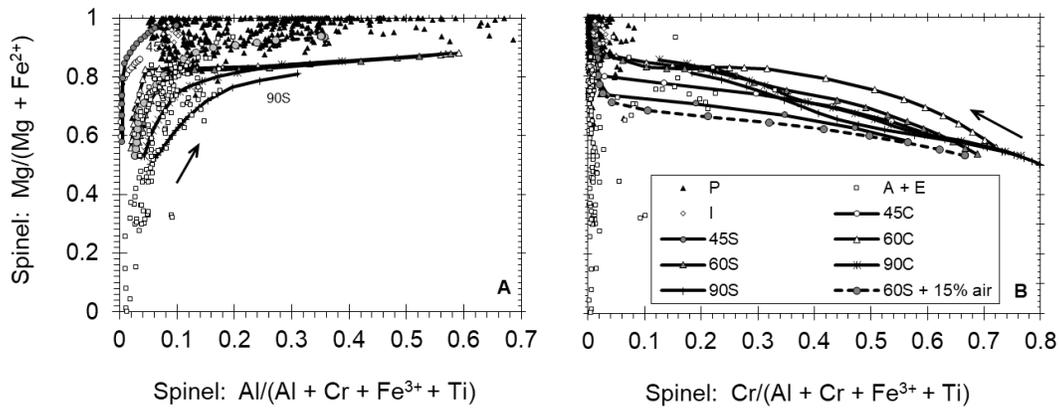

**Figure 2.** Measured atom compositions (without nickel) of Cretaceous-Paleogene (K-P) boundary spinel grains (Table DR1 [see footnote 1]), compared to calculated compositions along a representative pressure-temperature (*P-T*) trajectory (with temperature decreasing in the direction of arrow; see "x" curve in Fig. 1) of carbonate-rich (C) and sulfate-rich (S) plume compositions (Table 1). Sites: A + E—Atlantic Ocean and Europe; P—Pacific Ocean; I—Indian Ocean.

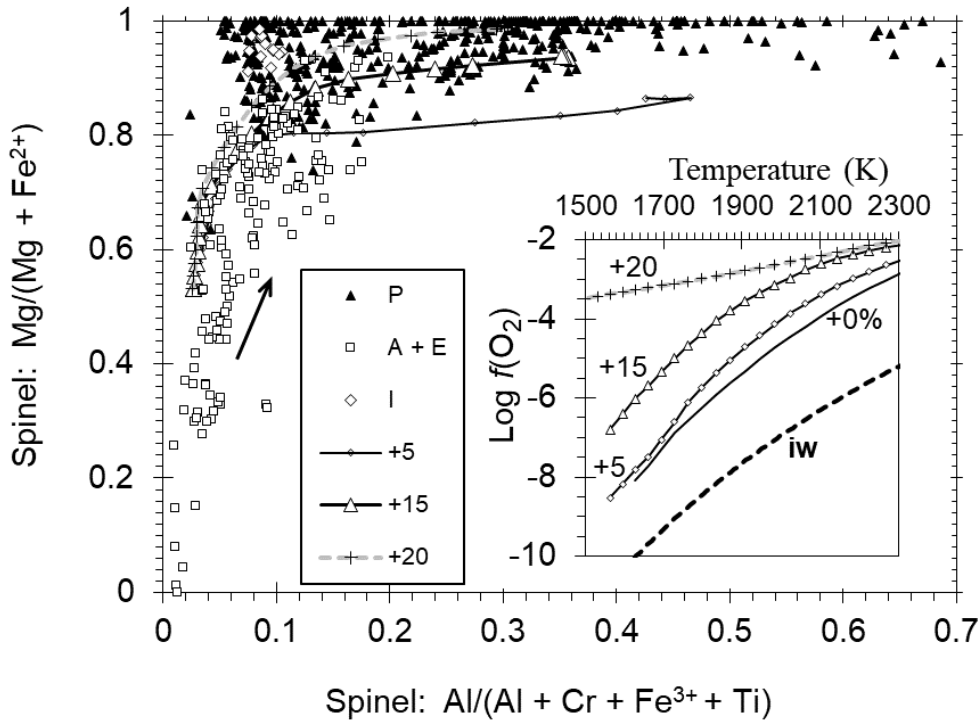

**Figure 3.** Effect of admixed air on calculated spinel compositions, as temperature decreases (arrow), for vapor plume 60S with 5%, 15%, and 20% admixed air. Inset shows corresponding oxygen fugacities as in Figure 1. Line marked iw shows iron-wüstite (Fe-FeO) oxygen-fugacity buffer curve. Spinel data as in Figure 2. Sites: A + E—Atlantic Ocean and Europe; P—Pacific Ocean; I—Indian Ocean.

Figure 3 illustrates the effect on spinel composition of adding successive increments of air to an initial vapor of composition 60S. Admixture of air progressively increases $f(O_2)$ of the vapor, shifting the path of spinel composition (Fig. 3) toward lower $Al/(Al + Cr + Fe^{3+} + Ti)$ and higher $Mg/(Mg + Fe^{2+})$. Composition 60S with 15 atom percent admixed air most closely matches the range of observed spinel compositions and passes as much as ~0.2 log units above the NNO buffer curve between 2000 and 1650 K. The chemical processes calculated to occur in this cooling, expanding vapor plume—if strict chemical equilibrium is assumed—begin near 2400 K, where ~45% of all the atoms in the plume are condensed, 99% as silicate liquid (27.3 wt% $SiO_2$, 0.2 $TiO_2$, 4.7 $Al_2O_3$, 2.4 $Fe_2O_3$, 12.8 FeO, 10.0 MgO, 42.7 CaO). Fe-rich spinel crystallizes from liquid in trace amounts and contains 64% of the Cr in the system while most Mg, Al, and Fe remain in the silicate liquid. The prediction of only trace Cr-rich spinel is consistent with its rarity in the K-P boundary layer (Fig. 2B). With decreasing temperature, spinel becomes much more abundant but poorer in Cr (Fig. 2B), and the $Fe^{3+}/Fe^{2+}$ ratio in spinel increases from ~1.3 at 2260 K to ~3.7 at 2100 K, and ~8.7 at 1908 K, both trends consistent with observed zoning (Preisinger et al., 1997). The $Fe^{3+}/Fe^{2+}$ ratio increases more steeply with decreasing temperature below 1900 K, reaching ~17 at 1695 K. At 2060 K, spinel contains ~14% of the Fe in the system, and 4% of all the atoms in the system are contained in spinel, coexisting with solid calcium silicate (primarily larnite) and liquid. As the vapor cools further, spinel abundance increases slowly, and its Mg and Al contents increase relative to Fe (Fig. 3). At 1908 K, spinel contains 80% of the Fe, 29% of the Mg, and 15% of the Al, and these amounts rise steadily to 82%, 44%, and 57%, respectively, at 1728 K, where the liquid composition is 32.1 wt% $SiO_2$, 0.004 $TiO_2$, 7.7 $Al_2O_3$, 6.5 $Fe_2O_3$, 2.6 FeO, 14.8 MgO, 34.2 CaO, 2% $Na_2O$. The remaining Mg crystallizes as periclase (MgO), followed by Ca-rich olivine (monticellite, approximately $CaMgSiO_4$) at 1628 K.

The calculation assumes continuous equilibration between vapor, liquid, spinel, and other solids. In the actual vapor plume, some liquid droplets would freeze at high temperatures, and the resulting compositions of their included spinels would reflect vapor conditions at the moment of freezing, yielding an array of compositions along the *P-T* path. It is expected that most liquid droplets containing crystalline spinel would quench to glass spherules above 1700 K, containing all the Ca, Al, Si, Mg, Fe, Ni, Cr, and PGE in the system. No Ca-sulfate or calcite is predicted to form as a primary condensate, but sulfate aerosols in the plume could react with primary condensates during cooling.

Alternative initial vapor compositions can be supposed that condense spinels quite inconsistent with those observed. If the proportions of sandstone and shale are substantially increased, very little spinel is produced in our simulations, and what is produced is Cr rich (chromite). Extremely projectile-rich, low-$f(O_2)$ vapor also condenses only small amounts of spinel, also rich in Cr, coexisting with Fe-rich metal. Spinels in Archean spherules have $Mg/(Mg + Fe^{2+} + Ni + Mn) < 0.22$, $Al/(Al + Cr + Fe^{3+} + Ti) < 0.12$, and $Cr/(Al + Cr + Fe^{3+} + Ti) > 0.54$ (Byerly and Lowe, 1994), consistent with our condensation calculations (not illustrated here) for an impactor-dominated vapor and/or a basaltic ocean-crust target. None of the predicted K-P vapor compositions allow condensation of $Fe^{2+}$-rich spinels (Fig. 2), so reducing conditions in some part of the vapor plume must have prevailed to produce the small subset of spinels with $Mg/(Mg + Fe^{2+}) < 0.5$ (Gerasimov, 2002).

## CONCLUSIONS

Once-glassy spherules marking the basal Ir-rich K-P boundary clay, and their included magnesioferrite spinels, all formed by condensation from the high-temperature vapor of the Chicxulub fireball. The carbonate + anhydrite target rock is crucial in establishing the high $f(O_2)$ of the plume; neither atmospheric entrainment nor ablation is required. The predicted condensates for 45° and 90° angle impacts bracket a large fraction of the observed spinel data. Predicted $f(O_2)$ values are consistent with the Ni and PGE contents of spinels. Variations in the observed compositions are the result of local physicochemical heterogeneities in the vapor plume, particularly in $f(O_2)$, and the effect of quenching spinel at different temperatures in the cooling spherules.

## ACKNOWLEDGMENTS

This work was supported by U.S. National Aeronautics and Space Administration grants NAG5-4476 and NAG5-11588 (to Grossman) and NAG5-12855 (to Ebel). We thank Frank Kyte for spinel data, expertise, and stimulating conversation.

Ebel & Grossman (2005) Chicxulub Spherule Condensation                                    p. 10

Ebel and Grossman (2005) *Geology*
Data Repository Table DR1:
**Table DR1: K/P spinel data, weight fractions of oxides (F. Kyte, 1998, personal commun.).**

| line | label | sample | FeO | Fe2O3 | SiO2 | MgO | Al2O3 | NiO | MnO | Cr2O3 | TiO2 | CaO | total |
|---|---|---|---|---|---|---|---|---|---|---|---|---|---|
| | **PACIFIC** | | | | | | | | | | | | |
| | DSDP 577 | | | | | | | | | | | | |
| 1 | 5 old | DSDP577 | 0.0000 | 0.6143 | 0.0020 | 0.2166 | 0.1602 | 0.0121 | 0.0072 | 0.0009 | 0.0032 | 0.0036 | 1.0201 |
| 2 | 6 old | DSDP577 | 0.0000 | 0.5783 | 0.0012 | 0.2111 | 0.1733 | 0.0107 | 0.0102 | 0.0005 | 0.0025 | 0.0021 | 0.9899 |
| 3 | 9 old | DSDP577 | 0.0051 | 0.5833 | 0.0020 | 0.2053 | 0.1999 | 0.0289 | 0.0046 | 0.0010 | 0.0025 | 0.0032 | 1.0358 |
| 4 | 10 old | DSDP577 | 0.0056 | 0.6189 | 0.0007 | 0.1940 | 0.1583 | 0.0263 | 0.0099 | 0.0002 | 0.0014 | 0.0040 | 1.0191 |
| 5 | 11 old | DSDP577 | 0.0022 | 0.6855 | 0.0012 | 0.1854 | 0.0877 | 0.0221 | 0.0123 | 0.0004 | 0.0015 | 0.0047 | 1.0030 |
| 6 | 12 old | DSDP577 | 0.0000 | 0.5893 | 0.0015 | 0.2067 | 0.1949 | 0.0283 | 0.0064 | 0.0007 | 0.0016 | 0.0039 | 1.0333 |
| 7 | 13 old | DSDP577 | 0.0000 | 0.5790 | 0.0010 | 0.2173 | 0.1876 | 0.0139 | 0.0040 | 0.0014 | 0.0040 | 0.0050 | 1.0132 |
| 8 | 14 old | DSDP577 | 0.0028 | 0.5696 | 0.0003 | 0.2104 | 0.2007 | 0.0153 | 0.0040 | 0.0006 | 0.0035 | 0.0053 | 1.0126 |
| 9 | 15 old | DSDP577 | 0.0000 | 0.5816 | 0.0003 | 0.2119 | 0.1889 | 0.0147 | 0.0036 | 0.0026 | 0.0030 | 0.0048 | 1.0114 |
| 10 | 16 old | DSDP577 | 0.0026 | 0.5441 | 0.0002 | 0.2140 | 0.2266 | 0.0139 | 0.0043 | 0.0005 | 0.0023 | 0.0050 | 1.0134 |
| 11 | 17 old | DSDP577 | 0.0000 | 0.6053 | 0.0002 | 0.2045 | 0.1573 | 0.0153 | 0.0045 | 0.0030 | 0.0030 | 0.0056 | 0.9989 |
| 12 | 18 old | DSDP577 | 0.0000 | 0.5465 | 0.0005 | 0.2120 | 0.2111 | 0.0121 | 0.0043 | 0.0011 | 0.0034 | 0.0058 | 0.9969 |
| 13 | 19 old | DSDP577 | 0.0058 | 0.5560 | 0.0002 | 0.2022 | 0.1914 | 0.0143 | 0.0051 | 0.0032 | 0.0041 | 0.0062 | 0.9886 |
| 14 | 12 old | DSDP577 | 0.0000 | 0.6107 | 0.0013 | 0.2116 | 0.1684 | 0.0149 | 0.0052 | 0.0018 | 0.0032 | 0.0079 | 1.0251 |
| 15 | 13 old | DSDP577 | 0.0000 | 0.6025 | 0.0006 | 0.2185 | 0.1789 | 0.0142 | 0.0053 | 0.0022 | 0.0035 | 0.0074 | 1.0331 |
| 16 | 15 old | DSDP577 | 0.0000 | 0.5795 | 0.0019 | 0.2072 | 0.1763 | 0.0063 | 0.0054 | 0.0010 | 0.0030 | 0.0082 | 0.9887 |
| 17 | 16 old | DSDP577 | 0.0060 | 0.5951 | 0.0005 | 0.1945 | 0.1558 | 0.0144 | 0.0057 | 0.0016 | 0.0034 | 0.0088 | 0.9859 |
| 18 | 17 old | DSDP577 | 0.0022 | 0.4158 | 0.0015 | 0.2250 | 0.3199 | 0.0027 | 0.0039 | 0.0005 | 0.0019 | 0.0046 | 0.9780 |
| 19 | 18 old | DSDP577 | 0.0000 | 0.4327 | 0.0007 | 0.2270 | 0.3033 | 0.0024 | 0.0045 | 0.0002 | 0.0037 | 0.0054 | 0.9798 |
| 20 | 20 old | DSDP577 | 0.0000 | 0.4839 | 0.0015 | 0.2209 | 0.2599 | 0.0098 | 0.0056 | 0.0028 | 0.0052 | 0.0073 | 0.9969 |
| 21 | 21 old | DSDP577 | 0.0000 | 0.4616 | 0.0016 | 0.2243 | 0.2793 | 0.0104 | 0.0054 | 0.0044 | 0.0044 | 0.0057 | 0.9970 |
| 22 | 22 old | DSDP577 | 0.0000 | 0.5274 | 0.0035 | 0.2039 | 0.2080 | 0.0182 | 0.0054 | 0.0014 | 0.0033 | 0.0050 | 0.9761 |
| 23 | 23 old | DSDP577 | 0.0060 | 0.5378 | 0.0008 | 0.2042 | 0.2163 | 0.0175 | 0.0060 | 0.0021 | 0.0018 | 0.0039 | 0.9964 |
| 24 | 24 old | DSDP577 | 0.0019 | 0.5108 | 0.0022 | 0.2094 | 0.2405 | 0.0211 | 0.0046 | 0.0020 | 0.0029 | 0.0035 | 0.9988 |
| 25 | 24 old | DSDP577 | 0.0000 | 0.5409 | 0.0007 | 0.2156 | 0.1980 | 0.0197 | 0.0037 | 0.0097 | 0.0041 | 0.0033 | 0.9956 |
| 26 | 25 old | DSDP577 | 0.0029 | 0.4775 | 0.0020 | 0.2175 | 0.2575 | 0.0068 | 0.0054 | 0.0008 | 0.0071 | 0.0048 | 0.9823 |
| 27 | 26 old | DSDP577 | 0.0000 | 0.4878 | 0.0014 | 0.2229 | 0.2415 | 0.0051 | 0.0062 | 0.0013 | 0.0065 | 0.0063 | 0.9788 |
| 28 | 29 old | DSDP577 | 0.0000 | 0.4646 | 0.0020 | 0.2371 | 0.2537 | 0.0062 | 0.0059 | 0.0013 | 0.0061 | 0.0047 | 0.9817 |
| 29 | 30 old | DSDP577 | 0.0251 | 0.5338 | 0.0007 | 0.1803 | 0.1924 | 0.0230 | 0.0045 | 0.0027 | 0.0014 | 0.0045 | 0.9684 |
| 30 | 5 old | DSDP577 | 0.0109 | 0.5770 | 0.0008 | 0.1903 | 0.1814 | 0.0270 | 0.0070 | 0.0011 | 0.0021 | 0.0060 | 1.0037 |
| 31 | 6 old | DSDP577 | 0.0145 | 0.5600 | 0.0006 | 0.1921 | 0.1962 | 0.0242 | 0.0068 | 0.0008 | 0.0036 | 0.0072 | 1.0060 |
| 32 | 7 old | DSDP577 | 0.0085 | 0.5673 | 0.0016 | 0.1932 | 0.1914 | 0.0282 | 0.0070 | 0.0011 | 0.0033 | 0.0069 | 1.0084 |
| 33 | 9 old | DSDP577 | 0.0000 | 0.5628 | 0.0014 | 0.2188 | 0.2010 | 0.0029 | 0.0045 | 0.0003 | 0.0042 | 0.0082 | 1.0041 |
| 34 | 10 old | DSDP577 | 0.0000 | 0.5674 | 0.0021 | 0.2147 | 0.1926 | 0.0088 | 0.0050 | 0.0012 | 0.0030 | 0.0081 | 1.0028 |
| 35 | 11 old | DSDP577 | 0.0000 | 0.5659 | 0.0013 | 0.2172 | 0.2091 | 0.0081 | 0.0052 | 0.0007 | 0.0025 | 0.0072 | 1.0172 |
| 36 | 12 old | DSDP577 | 0.0356 | 0.7008 | 0.0010 | 0.1242 | 0.0348 | 0.0578 | 0.0220 | 0.0014 | 0.0016 | 0.0059 | 0.9852 |
| 37 | 14 old | DSDP577 | 0.0202 | 0.4857 | 0.0013 | 0.2187 | 0.2858 | 0.0061 | 0.0054 | 0.0004 | 0.0037 | 0.0036 | 1.0309 |
| 38 | 15 old | DSDP577 | 0.0208 | 0.4939 | 0.0022 | 0.2118 | 0.2686 | 0.0041 | 0.0067 | 0.0004 | 0.0018 | 0.0041 | 1.0145 |
| 39 | 18 old | DSDP577 | 0.0027 | 0.3347 | 0.0010 | 0.2423 | 0.4132 | 0.0084 | 0.0029 | 0.0046 | 0.0027 | 0.0021 | 1.0144 |
| 40 | 19 old | DSDP577 | 0.0066 | 0.3525 | 0.0009 | 0.2336 | 0.3892 | 0.0085 | 0.0036 | 0.0013 | 0.0027 | 0.0024 | 1.0013 |
| 41 | 20 old | DSDP577 | 0.0092 | 0.3303 | 0.0009 | 0.2293 | 0.3896 | 0.0084 | 0.0031 | 0.0068 | 0.0030 | 0.0017 | 0.9825 |
| 42 | 21 old | DSDP577 | 0.0000 | 0.3188 | 0.0008 | 0.2475 | 0.4229 | 0.0076 | 0.0031 | 0.0048 | 0.0022 | 0.0023 | 1.0101 |
| 43 | 22 old | DSDP577 | 0.0110 | 0.3439 | 0.0017 | 0.2242 | 0.3768 | 0.0094 | 0.0040 | 0.0026 | 0.0031 | 0.0021 | 0.9789 |
| 44 | 23 old | DSDP577 | 0.0078 | 0.5940 | 0.0004 | 0.2044 | 0.1731 | 0.0106 | 0.0062 | 0.0035 | 0.0035 | 0.0062 | 1.0096 |
| 45 | 24 old | DSDP577 | 0.0175 | 0.5959 | 0.0011 | 0.1989 | 0.1727 | 0.0114 | 0.0057 | 0.0036 | 0.0032 | 0.0060 | 1.0160 |
| 46 | 26 old | DSDP577 | 0.0116 | 0.5987 | 0.0017 | 0.2016 | 0.1570 | 0.0096 | 0.0040 | 0.0082 | 0.0043 | 0.0052 | 1.0019 |
| 47 | 27 old | DSDP577 | 0.0194 | 0.5751 | 0.0024 | 0.2018 | 0.1823 | 0.0082 | 0.0044 | 0.0004 | 0.0042 | 0.0022 | 1.0003 |
| 48 | 28 old | DSDP577 | 0.0000 | 0.6016 | 0.0017 | 0.2131 | 0.1662 | 0.0071 | 0.0080 | 0.0000 | 0.0039 | 0.0015 | 1.0030 |
| 49 | 1 24viii93 | 577JB93-3,9 | 0.0041 | 0.5220 | 0.0004 | 0.2053 | 0.2268 | 0.0253 | 0.0034 | 0.0008 | 0.0039 | 0.0042 | 0.9963 |
| 50 | 2 24viii93 | 577JB93-3,9 | 0.0157 | 0.5432 | 0.0006 | 0.1959 | 0.2118 | 0.0276 | 0.0038 | 0.0007 | 0.0026 | 0.0038 | 1.0056 |
| 51 | 3 24viii93 | 577JB93-3,9 | 0.0164 | 0.5368 | 0.0006 | 0.1929 | 0.2123 | 0.0286 | 0.0043 | 0.0005 | 0.0022 | 0.0037 | 0.9982 |
| 52 | 4 24viii93 | 577JB93-3,9 | 0.0059 | 0.5034 | 0.0004 | 0.2127 | 0.2509 | 0.0203 | 0.0037 | 0.0028 | 0.0038 | 0.0033 | 1.0070 |
| 53 | 5 24viii93 | 577JB93-3,9 | 0.0088 | 0.4958 | 0.0003 | 0.2065 | 0.2481 | 0.0211 | 0.0040 | 0.0016 | 0.0033 | 0.0034 | 0.9929 |
| 54 | 9 24viii93 | 577JB93-3,8 | 0.0147 | 0.4262 | 0.0001 | 0.2141 | 0.3145 | 0.0210 | 0.0030 | 0.0035 | 0.0047 | 0.0034 | 1.0052 |
| 55 | 10 24viii93 | 577JB93-3,3 | 0.0000 | 0.4848 | 0.0020 | 0.2292 | 0.2610 | 0.0055 | 0.0056 | 0.0009 | 0.0051 | 0.0084 | 1.0025 |
| 56 | 11 24viii93 | 577JB93-3,4 | 0.0295 | 0.6250 | 0.0003 | 0.1832 | 0.1015 | 0.0090 | 0.0128 | 0.0001 | 0.0156 | 0.0030 | 0.9802 |
| 57 | 12 24viii93 | 577JB93-3,4 | 0.0548 | 0.6344 | 0.0002 | 0.1592 | 0.0852 | 0.0077 | 0.0224 | 0.0000 | 0.0152 | 0.0036 | 0.9828 |
| 58 | 13 24viii93 | 577JB93-3,4 | 0.0437 | 0.6882 | 0.0000 | 0.1602 | 0.0547 | 0.0151 | 0.0195 | 0.0006 | 0.0116 | 0.0046 | 0.9982 |
| 59 | 14 24viii93 | 577JB93-3,5 | 0.0000 | 0.4897 | 0.0003 | 0.2238 | 0.2587 | 0.0102 | 0.0055 | 0.0006 | 0.0053 | 0.0045 | 0.9987 |
| 60 | 15 24viii93 | 577JB93-3,5 | 0.0007 | 0.6410 | 0.0009 | 0.1962 | 0.1101 | 0.0101 | 0.0084 | 0.0000 | 0.0065 | 0.0088 | 0.9827 |
| 61 | 17 24viii93 | 577JB93-3,6 | 0.0041 | 0.6887 | 0.0010 | 0.1789 | 0.0644 | 0.0132 | 0.0166 | 0.0003 | 0.0032 | 0.0086 | 0.9789 |
| 62 | 47 24viii93 | 577JB93-5,3 | 0.0000 | 0.5891 | 0.0000 | 0.2156 | 0.1766 | 0.0083 | 0.0047 | 0.0040 | 0.0024 | 0.0084 | 1.0090 |
| 63 | 48 24viii93 | 577JB93-5,3 | 0.0008 | 0.5941 | 0.0005 | 0.2159 | 0.1795 | 0.0079 | 0.0045 | 0.0028 | 0.0039 | 0.0068 | 1.0167 |
| 64 | 49 24viii93 | 577JB93-5,4 | 0.0234 | 0.5486 | 0.0001 | 0.1972 | 0.2023 | 0.0170 | 0.0054 | 0.0109 | 0.0026 | 0.0046 | 1.0122 |
| 65 | 50 24viii93 | 577JB93-5,4 | 0.0315 | 0.5402 | 0.0011 | 0.1907 | 0.2017 | 0.0192 | 0.0044 | 0.0073 | 0.0029 | 0.0040 | 1.0029 |
| 66 | 51 24viii93 | 577JB93-5,5 | 0.0000 | 0.5360 | 0.0008 | 0.2252 | 0.2271 | 0.0078 | 0.0048 | 0.0007 | 0.0052 | 0.0059 | 1.0134 |
| 67 | 52 24viii93 | 577JB93-5,5 | 0.0000 | 0.5592 | 0.0008 | 0.2209 | 0.2010 | 0.0065 | 0.0055 | 0.0011 | 0.0040 | 0.0069 | 1.0059 |
| 68 | 54 24viii93 | 577JB93-5,6 | 0.0000 | 0.5730 | 0.0016 | 0.2196 | 0.1858 | 0.0055 | 0.0041 | 0.0011 | 0.0018 | 0.0126 | 1.0051 |
| 69 | 56 24viii93 | 577JB93-5,9 | 0.0000 | 0.5486 | 0.0004 | 0.2198 | 0.2039 | 0.0041 | 0.0053 | 0.0009 | 0.0053 | 0.0101 | 0.9984 |
| 70 | 58 24viii93 | 577JB93-5,10 | 0.0000 | 0.4805 | 0.0012 | 0.2289 | 0.2668 | 0.0017 | 0.0051 | 0.0040 | 0.0050 | 0.0077 | 1.0011 |
| 71 | 59 24viii93 | 577JB93-5,11 | 0.0000 | 0.5006 | 0.0002 | 0.2301 | 0.2660 | 0.0047 | 0.0035 | 0.0009 | 0.0026 | 0.0072 | 1.0159 |
| 72 | 60 24viii93 | 577JB93-5,11 | 0.0000 | 0.5603 | 0.0001 | 0.2211 | 0.2047 | 0.0058 | 0.0054 | 0.0004 | 0.0042 | 0.0112 | 1.0134 |
| 73 | 63 24viii93 | 577JB93-5,17 | 0.0034 | 0.5160 | 0.0003 | 0.2226 | 0.2567 | 0.0042 | 0.0025 | 0.0000 | 0.0008 | 0.0091 | 1.0156 |
| 74 | 64 24viii93 | 577JB93-5,17 | 0.0000 | 0.6062 | 0.0008 | 0.2154 | 0.1732 | 0.0066 | 0.0049 | 0.0011 | 0.0030 | 0.0102 | 1.0213 |
| 75 | 65 24viii93 | 577JB93-5,19 | 0.0185 | 0.5927 | 0.0002 | 0.1913 | 0.1764 | 0.0170 | 0.0095 | 0.0002 | 0.0012 | 0.0090 | 1.0159 |
| 76 | 66 24viii93 | 577JB93-5,19 | 0.0038 | 0.5982 | 0.0015 | 0.1924 | 0.1693 | 0.0347 | 0.0083 | 0.0000 | 0.0018 | 0.0079 | 1.0179 |
| 77 | 70 24viii93 | 577JB93-5,26a | 0.0000 | 0.5772 | 0.0000 | 0.2247 | 0.1959 | 0.0058 | 0.0031 | 0.0054 | 0.0030 | 0.0071 | 1.0221 |
| 78 | 71 24viii93 | 577JB93-5,26 | 0.0000 | 0.4648 | 0.0002 | 0.2359 | 0.2947 | 0.0092 | 0.0050 | 0.0009 | 0.0046 | 0.0043 | 1.0195 |
| 79 | 72 24viii93 | 577JB93-5,26 | 0.0000 | 0.4631 | 0.0000 | 0.2348 | 0.2932 | 0.0090 | 0.0044 | 0.0023 | 0.0033 | 0.0043 | 1.0143 |
| 80 | 73 24viii93 | 577JB93-5,25 | 0.0641 | 0.6912 | 0.0005 | 0.1631 | 0.0744 | 0.0070 | 0.0066 | 0.0005 | 0.0022 | 0.0023 | 1.0119 |
| 81 | 74 24viii93 | 577JB93-5,25 | 0.0433 | 0.6877 | 0.0006 | 0.1775 | 0.0782 | 0.0070 | 0.0065 | 0.0049 | 0.0025 | 0.0019 | 1.0101 |
| 82 | 75 24viii93 | 577JB93-5,24 | 0.0000 | 0.6804 | 0.0010 | 0.1960 | 0.0929 | 0.0337 | 0.0044 | 0.0004 | 0.0030 | 0.0039 | 1.0157 |
| 83 | 76 24viii93 | 577JB93-5,24 | 0.0000 | 0.6696 | 0.0005 | 0.2053 | 0.0984 | 0.0287 | 0.0050 | 0.0009 | 0.0030 | 0.0049 | 1.0163 |
| 84 | 83 24viii93 | 577JB93-5,45 | 0.0245 | 0.3419 | 0.0009 | 0.2211 | 0.4020 | 0.0209 | 0.0027 | 0.0025 | 0.0023 | 0.0026 | 1.0215 |
| 85 | 84 24viii93 | 577JB93-5,45 | 0.0303 | 0.2986 | 0.0017 | 0.2189 | 0.4271 | 0.0209 | 0.0029 | 0.0055 | 0.0020 | 0.0017 | 1.0096 |
| 86 | 85 24viii93 | 577JB93-5,42 | 0.0909 | 0.6726 | 0.0003 | 0.0984 | 0.0100 | 0.0588 | 0.0122 | 0.0445 | 0.0028 | 0.0011 | 0.9916 |
| 87 | 86 24viii93 | 577JB93-5,42 | 0.0861 | 0.7123 | 0.0015 | 0.1088 | 0.0121 | 0.0559 | 0.0079 | 0.0054 | 0.0042 | 0.0007 | 0.9949 |
| 88 | 87 24viii93 | 577JB93-5,40 | 0.0361 | 0.6673 | 0.0013 | 0.1665 | 0.0841 | 0.0184 | 0.0103 | 0.0030 | 0.0035 | 0.0072 | 0.9976 |

| | | | | | | | | | | | | | | |
|---|---|---|---|---|---|---|---|---|---|---|---|---|---|---|
| 89 | 88 | 24viii93 | 577JB93-5,40 | 0.0256 | 0.6881 | 0.0004 | 0.1782 | 0.0775 | 0.0107 | 0.0124 | 0.0014 | 0.0052 | 0.0072 | 1.0069 |
| 90 | 89 | 24viii93 | 577JB93-5,39 | 0.0000 | 0.5327 | 0.0002 | 0.2139 | 0.2277 | 0.0145 | 0.0048 | 0.0017 | 0.0025 | 0.0055 | 1.0034 |
| 91 | 90 | 24viii93 | 577JB93-5,39 | 0.0000 | 0.5409 | 0.0008 | 0.2216 | 0.2310 | 0.0139 | 0.0047 | 0.0017 | 0.0025 | 0.0060 | 1.0230 |
| 92 | 91 | 24viii93 | 577JB93-5,38 | 0.0036 | 0.6145 | 0.0000 | 0.2071 | 0.1651 | 0.0054 | 0.0052 | 0.0010 | 0.0035 | 0.0128 | 1.0182 |
| 93 | 93 | 24viii93 | 577JB93-5,35 | 0.0314 | 0.5721 | 0.0003 | 0.1982 | 0.1892 | 0.0034 | 0.0073 | 0.0000 | 0.0070 | 0.0068 | 1.0158 |
| 94 | 94 | 24viii93 | 577JB93-5,35 | 0.0444 | 0.5617 | 0.0047 | 0.1885 | 0.1742 | 0.0035 | 0.0081 | 0.0004 | 0.0081 | 0.0074 | 1.0010 |
| 95 | 97 | 24viii93 | 577JB93-5,31 | 0.0000 | 0.7033 | 0.0006 | 0.1919 | 0.0621 | 0.0123 | 0.0206 | 0.0001 | 0.0071 | 0.0036 | 1.0015 |
| 96 | 98 | 24viii93 | 577JB93-5,31 | 0.0185 | 0.6409 | 0.0018 | 0.1802 | 0.1233 | 0.0331 | 0.0090 | 0.0003 | 0.0033 | 0.0038 | 1.0143 |
| 97 | 99 | 24viii93 | 577JB93-5,31 | 0.0185 | 0.6637 | 0.0019 | 0.1740 | 0.0907 | 0.0292 | 0.0122 | 0.0002 | 0.0034 | 0.0033 | 0.9972 |
| 98 | 100 | 24viii93 | 577JB93-5,30 | 0.0393 | 0.7057 | 0.0006 | 0.1455 | 0.0505 | 0.0607 | 0.0058 | 0.0071 | 0.0037 | 0.0013 | 1.0202 |
| 99 | 101 | 24viii93 | 577JB93-5,30 | 0.0287 | 0.6882 | 0.0005 | 0.1572 | 0.0595 | 0.0490 | 0.0032 | 0.0133 | 0.0019 | 0.0026 | 1.0040 |
| 100 | 103 | 24viii93 | 577JB93-5,30 | 0.0319 | 0.7028 | 0.0005 | 0.1492 | 0.0491 | 0.0575 | 0.0042 | 0.0067 | 0.0026 | 0.0020 | 1.0065 |
| 101 | 104 | 24viii93 | 577JB93-5,30 | 0.0493 | 0.6881 | 0.0002 | 0.1347 | 0.0389 | 0.0544 | 0.0064 | 0.0083 | 0.0048 | 0.0014 | 0.9865 |
| 102 | 41 | 9ix93 | 577JB93-4-19 | 0.0005 | 0.4746 | 0.0018 | 0.2240 | 0.2580 | 0.0025 | 0.0058 | 0.0004 | 0.0121 | 0.0079 | 0.9874 |
| 103 | 42 | 9ix93 | 577JB93-4-19 | 0.0000 | 0.4523 | 0.0005 | 0.2436 | 0.3017 | 0.0015 | 0.0057 | 0.0000 | 0.0069 | 0.0038 | 1.0159 |
| 104 | 43 | 9ix93 | 577JB93-4-19 | 0.0000 | 0.4098 | 0.0015 | 0.2429 | 0.3408 | 0.0024 | 0.0050 | 0.0033 | 0.0074 | 0.0032 | 1.0162 |
| 105 | 44 | 9ix93 | 577JB93-4-18 | 0.0000 | 0.5515 | 0.0016 | 0.2240 | 0.1962 | 0.0046 | 0.0070 | 0.0059 | 0.0043 | 0.0087 | 1.0037 |
| 106 | 46 | 9ix93 | 577JB93-4-18 | 0.0000 | 0.4433 | 0.0004 | 0.2426 | 0.3120 | 0.0036 | 0.0043 | 0.0055 | 0.0032 | 0.0050 | 1.0199 |
| 107 | 48 | 9ix93 | 577JB93-4-17 | 0.0000 | 0.5953 | 0.0001 | 0.2191 | 0.1745 | 0.0042 | 0.0059 | 0.0010 | 0.0042 | 0.0102 | 1.0147 |
| 108 | 50 | 9ix93 | 577JB93-4-14 | 0.0000 | 0.6289 | 0.0001 | 0.2088 | 0.1490 | 0.0054 | 0.0046 | 0.0029 | 0.0005 | 0.0133 | 1.0135 |
| 109 | 51 | 9ix93 | 577JB93-4-14 | 0.0000 | 0.6117 | 0.0016 | 0.2234 | 0.1614 | 0.0051 | 0.0036 | 0.0007 | 0.0008 | 0.0101 | 1.0183 |
| 110 | 52 | 9ix93 | 577JB93-4-14 | 0.0000 | 0.6058 | 0.0012 | 0.2194 | 0.1667 | 0.0056 | 0.0039 | 0.0006 | 0.0008 | 0.0129 | 1.0169 |
| 111 | 53 | 9ix93 | 577JB93-4-13 | 0.0000 | 0.5343 | 0.0000 | 0.2239 | 0.2293 | 0.0045 | 0.0036 | 0.0000 | 0.0049 | 0.0078 | 1.0083 |
| 112 | 54 | 9ix93 | 577JB93-4-13 | 0.0000 | 0.5295 | 0.0015 | 0.2298 | 0.2295 | 0.0041 | 0.0040 | 0.0000 | 0.0049 | 0.0085 | 1.0116 |
| 113 | 55 | 9ix93 | 577JB93-4-13 | 0.0000 | 0.5764 | 0.0015 | 0.2206 | 0.1824 | 0.0053 | 0.0047 | 0.0025 | 0.0051 | 0.0092 | 1.0077 |
| 114 | 56 | 9ix93 | 577JB93-4-7 | 0.0000 | 0.6116 | 0.0015 | 0.2041 | 0.1492 | 0.0159 | 0.0093 | 0.0005 | 0.0066 | 0.0077 | 1.0064 |
| 115 | 57 | 9ix93 | 577JB93-4-7 | 0.0023 | 0.6234 | 0.0014 | 0.2003 | 0.1595 | 0.0189 | 0.0071 | 0.0012 | 0.0056 | 0.0082 | 1.0280 |
| 116 | 58 | 9ix93 | 577JB93-4-7 | 0.0000 | 0.6245 | 0.0010 | 0.2037 | 0.1468 | 0.0154 | 0.0099 | 0.0001 | 0.0056 | 0.0087 | 1.0158 |
| 117 | 60 | 9ix93 | 577JB93-4-8 | 0.0005 | 0.6097 | 0.0004 | 0.1961 | 0.1328 | 0.0282 | 0.0044 | 0.0312 | 0.0027 | 0.0036 | 1.0095 |
| 118 | 62 | 9ix93 | 577JB93-4-8 | 0.0065 | 0.6884 | 0.0010 | 0.1766 | 0.0905 | 0.0315 | 0.0032 | 0.0048 | 0.0037 | 0.0034 | 1.0095 |
| 119 | 72 | 9ix93 | 577JB93-4-16 | 0.0000 | 0.5787 | 0.0011 | 0.2181 | 0.1705 | 0.0104 | 0.0091 | 0.0022 | 0.0107 | 0.0083 | 1.0092 |
| 120 | 75 | 9ix93 | 577JB93-4-11 | 0.0000 | 0.6408 | 0.0010 | 0.2073 | 0.1329 | 0.0125 | 0.0077 | 0.0017 | 0.0038 | 0.0074 | 1.0151 |
| 121 | 76 | 9ix93 | 577JB93-4-4 | 0.0015 | 0.4853 | 0.0000 | 0.2160 | 0.2776 | 0.0133 | 0.0058 | 0.0027 | 0.0028 | 0.0033 | 1.0083 |
| 122 | 77 | 9ix93 | 577JB93-4-4 | 0.0002 | 0.3942 | 0.0004 | 0.2322 | 0.3596 | 0.0148 | 0.0030 | 0.0024 | 0.0025 | 0.0027 | 1.0120 |
| 123 | 79 | 9ix93 | 577JB93-4-4 | 0.0009 | 0.4124 | 0.0006 | 0.2261 | 0.3424 | 0.0136 | 0.0037 | 0.0005 | 0.0030 | 0.0031 | 1.0065 |
| 124 | 81 | 9ix93 | 577JB93-4-9 | 0.0000 | 0.4311 | 0.0012 | 0.2314 | 0.3131 | 0.0083 | 0.0055 | 0.0024 | 0.0042 | 0.0040 | 1.0013 |
| 125 | 82 | 9ix93 | 577JB93-4-9 | 0.0043 | 0.7082 | 0.0004 | 0.0936 | 0.0257 | 0.1314 | 0.0233 | 0.0017 | 0.0013 | 0.0044 | 0.9942 |
| 126 | 83 | 9ix93 | 577JB93-4-9 | 0.0062 | 0.7328 | 0.0007 | 0.1284 | 0.0317 | 0.0763 | 0.0217 | 0.0016 | 0.0018 | 0.0054 | 1.0065 |
| 127 | DSDP 576 | | | | | | | | | | | | | |
| 128 | 1 | 30ix93 | 576JB93-4-1 | 0.0135 | 0.6317 | 0.0004 | 0.1979 | 0.1353 | 0.0100 | 0.0048 | 0.0104 | 0.0015 | 0.0058 | 1.0112 |
| 129 | 2 | 30ix93 | 576JB93-4-1 | 0.1058 | 0.7113 | 0.0001 | 0.1127 | 0.0153 | 0.0244 | 0.0094 | 0.0060 | 0.0043 | 0.0014 | 0.9908 |
| 130 | 3 | 30ix93 | 576JB93-4-1 | 0.1129 | 0.7070 | 0.0016 | 0.1105 | 0.0198 | 0.0252 | 0.0098 | 0.0074 | 0.0036 | 0.0016 | 0.9995 |
| 131 | 6 | 30ix93 | 576JB93-5-11 | 0.0000 | 0.6209 | 0.0001 | 0.2075 | 0.1505 | 0.0136 | 0.0056 | 0.0014 | 0.0026 | 0.0079 | 1.0102 |
| 132 | 7 | 30ix93 | 576JB93-5-11 | 0.0015 | 0.5827 | 0.0000 | 0.2082 | 0.1806 | 0.0096 | 0.0048 | 0.0014 | 0.0031 | 0.0070 | 0.9989 |
| 133 | 8 | 30ix93 | 576JB93-5-10 | 0.0000 | 0.5845 | 0.0000 | 0.2102 | 0.1757 | 0.0083 | 0.0043 | 0.0024 | 0.0053 | 0.0098 | 1.0005 |
| 134 | 9 | 30ix93 | 576JB93-5-10 | 0.0000 | 0.5795 | 0.0000 | 0.2157 | 0.1755 | 0.0085 | 0.0042 | 0.0012 | 0.0058 | 0.0110 | 1.0013 |
| 135 | 10 | 30ix93 | 576JB93-5-12 | 0.0000 | 0.7379 | 0.0000 | 0.1929 | 0.0459 | 0.0094 | 0.0032 | 0.0090 | 0.0012 | 0.0114 | 1.0109 |
| 136 | 11 | 30ix93 | 576JB93-5-12 | 0.0000 | 0.7385 | 0.0000 | 0.1976 | 0.0445 | 0.0089 | 0.0040 | 0.0037 | 0.0019 | 0.0120 | 1.0112 |

| | | | | | | | | | | | | | |
|---|---|---|---|---|---|---|---|---|---|---|---|---|---|
| 137 | 12 | 30ix93 | 576JB93-5-23 | 0.0127 | 0.4418 | 0.0002 | 0.2229 | 0.3097 | 0.0137 | 0.0040 | 0.0060 | 0.0062 | 0.0034 | 1.0205 |
| 138 | 13 | 30ix93 | 576JB93-5-23 | 0.0248 | 0.3914 | 0.0000 | 0.2141 | 0.3330 | 0.0117 | 0.0036 | 0.0106 | 0.0049 | 0.0026 | 0.9967 |
| 139 | 14 | 30ix93 | 576JB93-5-23 | 0.0316 | 0.3827 | 0.0000 | 0.2108 | 0.3424 | 0.0113 | 0.0029 | 0.0083 | 0.0045 | 0.0035 | 0.9979 |
| 140 | 16 | 30ix93 | 576JB93-5-18 | 0.0114 | 0.5607 | 0.0000 | 0.2000 | 0.1928 | 0.0189 | 0.0042 | 0.0043 | 0.0025 | 0.0034 | 0.9982 |
| 141 | 17 | 30ix93 | 576JB93-5-18 | 0.0095 | 0.5755 | 0.0000 | 0.2068 | 0.1976 | 0.0167 | 0.0058 | 0.0042 | 0.0026 | 0.0037 | 1.0223 |
| 142 | 48 | 30ix93 | 576JB93-7-1 | 0.0080 | 0.5911 | 0.0004 | 0.2080 | 0.1740 | 0.0109 | 0.0032 | 0.0031 | 0.0037 | 0.0040 | 1.0063 |
| 143 | 49 | 30ix93 | 576JB93-7-1 | 0.0000 | 0.5906 | 0.0000 | 0.2190 | 0.1833 | 0.0085 | 0.0065 | 0.0056 | 0.0019 | 0.0046 | 1.0199 |
| 144 | 52 | 30ix93 | 576JB93-7-3 | 0.0226 | 0.6582 | 0.0003 | 0.1838 | 0.1130 | 0.0277 | 0.0033 | 0.0019 | 0.0036 | 0.0027 | 1.0171 |
| 145 | 53 | 30ix93 | 576JB93-7-3 | 0.0418 | 0.6583 | 0.0008 | 0.1635 | 0.0836 | 0.0281 | 0.0031 | 0.0044 | 0.0043 | 0.0025 | 0.9905 |
| 146 | 54 | 30ix93 | 576JB93-7-5 | 0.0230 | 0.5590 | 0.0000 | 0.1981 | 0.2023 | 0.0117 | 0.0047 | 0.0050 | 0.0018 | 0.0059 | 1.0116 |
| 147 | 55 | 30ix93 | 576JB93-7-5 | 0.0182 | 0.5628 | 0.0000 | 0.1998 | 0.1905 | 0.0047 | 0.0062 | 0.0033 | 0.0019 | 0.0058 | 0.9931 |
| 148 | 56 | 30ix93 | 576JB93-7-4 | 0.0385 | 0.6613 | 0.0004 | 0.1678 | 0.0877 | 0.0104 | 0.0048 | 0.0034 | 0.0028 | 0.0114 | 0.9885 |
| 149 | 57 | 30ix93 | 576JB93-7-4 | 0.0179 | 0.6290 | 0.0004 | 0.1876 | 0.1259 | 0.0041 | 0.0049 | 0.0009 | 0.0036 | 0.0144 | 0.9889 |
| 150 | 58 | 30ix93 | 576JB93-6-2 | 0.0000 | 0.6523 | 0.0017 | 0.1991 | 0.0834 | 0.0290 | 0.0026 | 0.0162 | 0.0027 | 0.0032 | 0.9902 |
| 151 | 59 | 30ix93 | 576JB93-6-2 | 0.0000 | 0.6279 | 0.0013 | 0.2235 | 0.1299 | 0.0167 | 0.0053 | 0.0019 | 0.0041 | 0.0038 | 1.0145 |
| 152 | 60 | 30ix93 | 576JB93-6-3 | 0.0000 | 0.6342 | 0.0000 | 0.2104 | 0.1425 | 0.0068 | 0.0057 | 0.0014 | 0.0034 | 0.0124 | 1.0167 |
| 153 | 61 | 30ix93 | 576JB93-6-3 | 0.0000 | 0.6383 | 0.0000 | 0.2070 | 0.1358 | 0.0068 | 0.0047 | 0.0017 | 0.0029 | 0.0120 | 1.0092 |
| 154 | 62 | 30ix93 | 576JB93-6-4 | 0.0175 | 0.5081 | 0.0003 | 0.2103 | 0.2543 | 0.0098 | 0.0052 | 0.0007 | 0.0030 | 0.0057 | 1.0150 |
| 155 | 64 | 30ix93 | 576JB93-6-5 | 0.0000 | 0.5864 | 0.0004 | 0.2153 | 0.1652 | 0.0061 | 0.0072 | 0.0100 | 0.0020 | 0.0046 | 0.9972 |
| 156 | 65 | 30ix93 | 576JB93-6-5 | 0.0370 | 0.5645 | 0.0004 | 0.1837 | 0.1774 | 0.0071 | 0.0073 | 0.0034 | 0.0016 | 0.0046 | 0.9871 |
| 157 | 66 | 30ix93 | 576JB93-6-6 | 0.0119 | 0.6100 | 0.0000 | 0.1943 | 0.1518 | 0.0244 | 0.0031 | 0.0019 | 0.0041 | 0.0038 | 1.0053 |
| 158 | 67 | 30ix93 | 576JB93-6-6 | 0.0000 | 0.5896 | 0.0005 | 0.2054 | 0.1706 | 0.0255 | 0.0032 | 0.0019 | 0.0042 | 0.0029 | 1.0037 |
| 159 | 68 | 30ix93 | 576JB93-6-7 | 0.0000 | 0.6494 | 0.0003 | 0.2038 | 0.1226 | 0.0070 | 0.0041 | 0.0018 | 0.0025 | 0.0101 | 1.0016 |
| 160 | 69 | 30ix93 | 576JB93-6-7 | 0.0000 | 0.6292 | 0.0001 | 0.2185 | 0.1393 | 0.0072 | 0.0044 | 0.0018 | 0.0025 | 0.0093 | 1.0122 |
| 161 | 70 | 30ix93 | 576JB93-6-8 | 0.0236 | 0.5884 | 0.0008 | 0.1920 | 0.1630 | 0.0111 | 0.0072 | 0.0002 | 0.0034 | 0.0030 | 0.9926 |
| 162 | 71 | 30ix93 | 576JB93-6-8 | 0.0094 | 0.5923 | 0.0007 | 0.2004 | 0.1642 | 0.0189 | 0.0037 | 0.0007 | 0.0048 | 0.0034 | 0.9986 |
| 163 | 72 | 30ix93 | 576JB93-6-9 | 0.0211 | 0.6963 | 0.0007 | 0.1863 | 0.0696 | 0.0137 | 0.0050 | 0.0087 | 0.0037 | 0.0027 | 1.0078 |
| 164 | 73 | 30ix93 | 576JB93-6-9 | 0.0159 | 0.7089 | 0.0006 | 0.1895 | 0.0632 | 0.0144 | 0.0036 | 0.0079 | 0.0031 | 0.0024 | 1.0094 |
| 165 | GPC3 | | | | | | | | | | | | | |
| 166 | 6 | old | GPC3 | 0.0716 | 0.6548 | 0.0009 | 0.1488 | 0.0867 | 0.0119 | 0.0075 | 0.0011 | 0.0045 | 0.0058 | 0.9935 |
| 167 | 7 | old | GPC3 | 0.0327 | 0.6034 | 0.0010 | 0.1847 | 0.1493 | 0.0123 | 0.0057 | 0.0026 | 0.0039 | 0.0037 | 0.9993 |
| 168 | 8 | old | GPC3 | 0.0499 | 0.6506 | 0.0001 | 0.1651 | 0.0957 | 0.0118 | 0.0067 | 0.0038 | 0.0042 | 0.0050 | 0.9928 |
| 169 | 9 | old | GPC3 | 0.0576 | 0.6244 | 0.0000 | 0.1620 | 0.1162 | 0.0109 | 0.0078 | 0.0024 | 0.0040 | 0.0044 | 0.9897 |
| 170 | 10 | old | GPC3 | 0.0203 | 0.5795 | 0.0044 | 0.1963 | 0.1522 | 0.0166 | 0.0084 | 0.0098 | 0.0161 | 0.0048 | 1.0085 |
| 171 | 11 | old | GPC3 | 0.0050 | 0.5826 | 0.0053 | 0.2118 | 0.1740 | 0.0120 | 0.0064 | 0.0064 | 0.0124 | 0.0068 | 1.0227 |
| 172 | 12 | old | GPC3 | 0.0113 | 0.5151 | 0.0035 | 0.2091 | 0.2188 | 0.0101 | 0.0071 | 0.0024 | 0.0115 | 0.0048 | 0.9937 |
| 173 | 14 | old | GPC3 | 0.0000 | 0.4956 | 0.0078 | 0.2330 | 0.2730 | 0.0063 | 0.0059 | 0.0021 | 0.0036 | 0.0084 | 1.0357 |
| 174 | 6 | old | GPC3 | 0.0427 | 0.7140 | 0.0056 | 0.1222 | 0.0112 | 0.0248 | 0.0377 | 0.0003 | 0.0019 | 0.0073 | 0.9677 |
| 175 | 7 | old | GPC3 | 0.0025 | 0.6127 | 0.0046 | 0.1992 | 0.1494 | 0.0177 | 0.0089 | 0.0007 | 0.0095 | 0.0108 | 1.0161 |
| 176 | 9 | old | GPC3 | 0.0414 | 0.6248 | 0.0038 | 0.1733 | 0.1132 | 0.0151 | 0.0037 | 0.0081 | 0.0047 | 0.0030 | 0.9911 |
| 177 | 19 | old | GPC3 | 0.0000 | 0.5480 | 0.0094 | 0.2190 | 0.2065 | 0.0112 | 0.0072 | 0.0031 | 0.0137 | 0.0072 | 1.0252 |
| 178 | 20 | old | GPC3 | 0.0060 | 0.5781 | 0.0049 | 0.2083 | 0.1672 | 0.0111 | 0.0071 | 0.0046 | 0.0143 | 0.0077 | 1.0094 |
| 179 | 20 | 24viii93 | GPC3JB93-8,4 | 0.0247 | 0.7064 | 0.0031 | 0.1747 | 0.0369 | 0.0216 | 0.0070 | 0.0053 | 0.0077 | 0.0027 | 0.9900 |
| 180 | 22 | 24viii93 | GPC3JB93-8,4 | 0.0017 | 0.7239 | 0.0003 | 0.1853 | 0.0283 | 0.0228 | 0.0085 | 0.0090 | 0.0073 | 0.0011 | 0.9882 |
| 181 | 24 | 24viii93 | GPC3JB93-8,5 | 0.0000 | 0.6025 | 0.0006 | 0.2191 | 0.1653 | 0.0150 | 0.0044 | 0.0096 | 0.0022 | 0.0064 | 1.0249 |
| 182 | 25 | 24viii93 | GPC3JB93-8,7 | 0.0053 | 0.6639 | 0.0004 | 0.1922 | 0.1003 | 0.0080 | 0.0045 | 0.0022 | 0.0025 | 0.0122 | 0.9916 |
| 183 | 26 | 24viii93 | GPC3JB93-8,8 | 0.0000 | 0.6635 | 0.0003 | 0.2015 | 0.0869 | 0.0142 | 0.0065 | 0.0051 | 0.0047 | 0.0039 | 0.9865 |
| 184 | 27 | 24viii93 | GPC3JB93-8,8 | 0.0000 | 0.6827 | 0.0004 | 0.2023 | 0.0872 | 0.0172 | 0.0041 | 0.0065 | 0.0038 | 0.0030 | 1.0071 |

| # | | | | | | | | | | | | | | |
|---|---|---|---|---|---|---|---|---|---|---|---|---|---|---|
| 185 | 28 | 24viii93 | GPC3JB93-8,9 | 0.0000 | 0.5186 | 0.0002 | 0.2300 | 0.2363 | 0.0083 | 0.0041 | 0.0013 | 0.0033 | 0.0070 | 1.0092 |
| 186 | 30 | 24viii93 | GPC3JB93-8,9 | 0.0002 | 0.5224 | 0.0007 | 0.2131 | 0.2236 | 0.0085 | 0.0037 | 0.0011 | 0.0034 | 0.0071 | 0.9837 |
| 187 | 36 | 24viii93 | GPC3JB93-6,5 | 0.0147 | 0.6548 | 0.0007 | 0.1990 | 0.1217 | 0.0129 | 0.0045 | 0.0040 | 0.0032 | 0.0048 | 1.0204 |
| 188 | 38 | 24viii93 | GPC3JB93-6,5 | 0.0000 | 0.6375 | 0.0061 | 0.1999 | 0.1209 | 0.0099 | 0.0062 | 0.0008 | 0.0023 | 0.0267 | 1.0103 |
| 189 | 42 | 24viii93 | GPC3JB93-6,10 | 0.0000 | 0.6941 | 0.0011 | 0.1972 | 0.0521 | 0.0248 | 0.0098 | 0.0075 | 0.0067 | 0.0030 | 0.9964 |
| 190 | 43 | 24viii93 | GPC3JB93-6,10 | 0.0000 | 0.6945 | 0.0019 | 0.1970 | 0.0449 | 0.0236 | 0.0090 | 0.0113 | 0.0067 | 0.0022 | 0.9911 |
| 191 | 45 | 24viii93 | GPC3JB93-6,2 | 0.0000 | 0.7151 | 0.0004 | 0.1922 | 0.0447 | 0.0164 | 0.0042 | 0.0127 | 0.0015 | 0.0068 | 0.9940 |
| 192 | 21 | 9ix93 | GPC3JB93-5-8 | 0.0016 | 0.6721 | 0.0019 | 0.1693 | 0.0484 | 0.0376 | 0.0057 | 0.0323 | 0.0031 | 0.0080 | 0.9801 |
| 193 | 23 | 9ix93 | GPC3JB93-7-2 | 0.0000 | 0.5151 | 0.0012 | 0.2163 | 0.2259 | 0.0175 | 0.0058 | 0.0013 | 0.0033 | 0.0078 | 0.9942 |
| 194 | 26 | 9ix93 | GPC3JB93-7-3 | 0.0000 | 0.5475 | 0.0003 | 0.2184 | 0.2006 | 0.0057 | 0.0052 | 0.0015 | 0.0032 | 0.0083 | 0.9907 |
| 195 | 27 | 9ix93 | GPC3JB93-7-3 | 0.0000 | 0.5787 | 0.0003 | 0.2129 | 0.1857 | 0.0074 | 0.0054 | 0.0026 | 0.0032 | 0.0075 | 1.0036 |
| 196 DSDP 886C | | | | | | | | | | | | | | |
| 197 | 4 | 27i94 | 886CJB94-1-2 | 0.0000 | 0.5575 | 0.0000 | 0.2181 | 0.2070 | 0.0123 | 0.0043 | 0.0087 | 0.0025 | 0.0058 | 1.0162 |
| 198 | 5 | 27i94 | 886CJB94-1-2 | 0.0000 | 0.5550 | 0.0003 | 0.2156 | 0.2067 | 0.0112 | 0.0043 | 0.0092 | 0.0032 | 0.0060 | 1.0115 |
| 199 | 6 | 27i94 | 886CJB94-1-2 | 0.0000 | 0.5422 | 0.0000 | 0.2208 | 0.2205 | 0.0089 | 0.0045 | 0.0023 | 0.0036 | 0.0063 | 1.0090 |
| 200 | 8 | 27i94 | 886CJB94-1-3 | 0.0152 | 0.6300 | 0.0000 | 0.1922 | 0.1307 | 0.0098 | 0.0042 | 0.0030 | 0.0034 | 0.0091 | 0.9976 |
| 201 | 9 | 27i94 | 886CJB94-1-3 | 0.0173 | 0.6397 | 0.0000 | 0.1915 | 0.1275 | 0.0093 | 0.0051 | 0.0022 | 0.0030 | 0.0087 | 1.0044 |
| 202 | 10 | 27i94 | 886CJB94-1-5 | 0.0000 | 0.5738 | 0.0002 | 0.2219 | 0.1953 | 0.0063 | 0.0042 | 0.0011 | 0.0037 | 0.0098 | 1.0164 |
| 203 | 11 | 27i94 | 886CJB94-1-5 | 0.0000 | 0.5757 | 0.0006 | 0.2111 | 0.1873 | 0.0060 | 0.0051 | 0.0010 | 0.0034 | 0.0098 | 0.9999 |
| 204 | 12 | 27i94 | 886CJB94-1-5 | 0.0005 | 0.5783 | 0.0003 | 0.2070 | 0.1756 | 0.0049 | 0.0045 | 0.0014 | 0.0029 | 0.0092 | 0.9845 |
| 205 | 13 | 27i94 | 886CJB94-1-6 | 0.0527 | 0.6709 | 0.0009 | 0.1445 | 0.0423 | 0.0429 | 0.0031 | 0.0382 | 0.0028 | 0.0018 | 1.0000 |
| 206 | 14 | 27i94 | 886CJB94-1-6 | 0.0548 | 0.6577 | 0.0006 | 0.1431 | 0.0438 | 0.0406 | 0.0027 | 0.0420 | 0.0034 | 0.0020 | 0.9907 |
| 207 | 15 | 27i94 | 886CJB94-1-6 | 0.0608 | 0.6917 | 0.0014 | 0.1398 | 0.0425 | 0.0399 | 0.0026 | 0.0088 | 0.0033 | 0.0028 | 0.9935 |
| 208 | 29 | 27i94 | 886CJB94-2-4 | 0.0000 | 0.5687 | 0.0000 | 0.2215 | 0.2080 | 0.0099 | 0.0057 | 0.0076 | 0.0015 | 0.0047 | 1.0274 |
| 209 | 30 | 27i94 | 886CJB94-2-4 | 0.0050 | 0.5577 | 0.0000 | 0.2126 | 0.2111 | 0.0109 | 0.0050 | 0.0079 | 0.0014 | 0.0049 | 1.0166 |
| 210 | 33 | 27i94 | 886CJB94-2-5 | 0.0000 | 0.6307 | 0.0000 | 0.2132 | 0.1502 | 0.0044 | 0.0041 | 0.0014 | 0.0051 | 0.0107 | 1.0199 |
| 211 | 36 | 27i94 | 886CJB94-3-11 | 0.0880 | 0.6689 | 0.0008 | 0.1395 | 0.0670 | 0.0217 | 0.0017 | 0.0159 | 0.0034 | 0.0036 | 1.0105 |
| 212 | 37 | 27i94 | 886CJB94-3-11 | 0.0797 | 0.6746 | 0.0009 | 0.1412 | 0.0568 | 0.0234 | 0.0021 | 0.0170 | 0.0035 | 0.0033 | 1.0024 |
| 213 | 38 | 27i94 | 886CJB94-3-11 | 0.0538 | 0.6692 | 0.0079 | 0.1631 | 0.0647 | 0.0241 | 0.0019 | 0.0163 | 0.0033 | 0.0077 | 1.0119 |
| 214 | 40 | 27i94 | 886CJB94-3-11A | 0.0472 | 0.6804 | 0.0000 | 0.1612 | 0.0681 | 0.0151 | 0.0085 | 0.0013 | 0.0055 | 0.0055 | 0.9928 |
| 215 | 41 | 27i94 | 886CJB94-3-11A | 0.0536 | 0.6903 | 0.0002 | 0.1606 | 0.0691 | 0.0159 | 0.0082 | 0.0008 | 0.0055 | 0.0052 | 1.0094 |
| 216 | 42 | 27i94 | 886CJB94-3-24 | 0.0350 | 0.6935 | 0.0015 | 0.1568 | 0.0442 | 0.0162 | 0.0036 | 0.0202 | 0.0018 | 0.0202 | 0.9931 |
| 217 | 43 | 27i94 | 886CJB94-3-24 | 0.0399 | 0.6992 | 0.0007 | 0.1542 | 0.0461 | 0.0170 | 0.0026 | 0.0136 | 0.0021 | 0.0196 | 0.9950 |
| 218 | 44 | 27i94 | 886CJB94-3-24 | 0.0427 | 0.7016 | 0.0012 | 0.1508 | 0.0423 | 0.0140 | 0.0019 | 0.0070 | 0.0015 | 0.0215 | 0.9847 |
| 219 | 45 | 27i94 | 886CJB94-3-26 | 0.0422 | 0.6533 | 0.0011 | 0.1796 | 0.0856 | 0.0119 | 0.0017 | 0.0296 | 0.0041 | 0.0022 | 1.0112 |
| 220 | 46 | 27i94 | 886CJB94-3-26 | 0.0449 | 0.6450 | 0.0003 | 0.1748 | 0.0859 | 0.0119 | 0.0022 | 0.0318 | 0.0033 | 0.0019 | 1.0019 |
| 221 DSDP 596 | | | | | | | | | | | | | | |
| 222 | 97 | 9 | 1,10 | 0.0163 | 0.6246 | 0.0000 | 0.1790 | 0.1259 | 0.0323 | 0.0032 | 0.0023 | 0.0021 | 0.0040 | 0.9897 |
| 223 | 107 | 19 | 1,11 | 0.0062 | 0.5851 | 0.0000 | 0.1892 | 0.1433 | 0.0299 | 0.0036 | 0.0310 | 0.0024 | 0.0058 | 0.9965 |
| 224 | 108 | 20 | 1,11 | 0.0126 | 0.5510 | 0.0001 | 0.1880 | 0.1441 | 0.0290 | 0.0039 | 0.0663 | 0.0021 | 0.0041 | 1.0012 |
| 225 | 109 | 21 | 1,11 | 0.0245 | 0.5660 | 0.0009 | 0.1789 | 0.1793 | 0.0371 | 0.0045 | 0.0077 | 0.0017 | 0.0047 | 1.0051 |
| 226 | 110 | 22 | 1,11 | 0.0103 | 0.5575 | 0.0001 | 0.1885 | 0.1447 | 0.0290 | 0.0037 | 0.0573 | 0.0019 | 0.0045 | 0.9977 |
| 227 | 2 | | 1,13 | 0.0078 | 0.6820 | 0.0005 | 0.1892 | 0.0780 | 0.0219 | 0.0017 | 0.0076 | 0.0023 | 0.0023 | 0.9933 |
| 228 | 5 | | 1,13 | 0.0289 | 0.6981 | 0.0008 | 0.1823 | 0.0728 | 0.0191 | 0.0048 | 0.0043 | 0.0054 | 0.0016 | 1.0182 |
| 229 | 7 | | 1,14 | 0.0087 | 0.6071 | 0.0004 | 0.1915 | 0.1357 | 0.0140 | 0.0040 | 0.0056 | 0.0028 | 0.0078 | 0.9776 |
| 230 | 8 | | 1,15 | 0.0057 | 0.7438 | 0.0006 | 0.1862 | 0.0376 | 0.0284 | 0.0040 | 0.0124 | 0.0012 | 0.0015 | 1.0215 |
| 231 | 9 | | 1,15 | 0.0025 | 0.7265 | 0.0011 | 0.1841 | 0.0377 | 0.0277 | 0.0044 | 0.0101 | 0.0016 | 0.0017 | 0.9974 |
| 232 | 11 | | 1,16 | 0.0036 | 0.6792 | 0.0008 | 0.1914 | 0.0813 | 0.0185 | 0.0055 | 0.0018 | 0.0027 | 0.0017 | 0.9865 |

| | | | | | | | | | | | | | |
|---|---|---|---|---|---|---|---|---|---|---|---|---|---|
| 233 | 13 | | 1,17 | 0.0322 | 0.6426 | 0.0003 | 0.1749 | 0.1145 | 0.0200 | 0.0060 | 0.0013 | 0.0017 | 0.0041 | 0.9977 |
| 234 | 14 | | 1,17 | 0.0329 | 0.6567 | 0.0000 | 0.1724 | 0.1080 | 0.0221 | 0.0078 | 0.0018 | 0.0015 | 0.0045 | 1.0075 |
| 235 | 15 | | 1,17 | 0.0207 | 0.6340 | 0.0005 | 0.1755 | 0.1191 | 0.0349 | 0.0034 | 0.0019 | 0.0023 | 0.0039 | 0.9961 |
| 236 | 16 | | 1,18 | 0.0191 | 0.6863 | 0.0007 | 0.1691 | 0.0317 | 0.0290 | 0.0038 | 0.0337 | 0.0026 | 0.0010 | 0.9769 |
| 237 | 17 | | 1,18 | 0.0218 | 0.7190 | 0.0013 | 0.1686 | 0.0384 | 0.0334 | 0.0049 | 0.0054 | 0.0028 | 0.0014 | 0.9969 |
| 238 | 18 | | 1,19 | 0.0379 | 0.6621 | 0.0008 | 0.1752 | 0.1027 | 0.0203 | 0.0076 | 0.0014 | 0.0044 | 0.0027 | 1.0151 |
| 239 | 19 | | 1,19 | 0.0273 | 0.6634 | 0.0003 | 0.1751 | 0.0963 | 0.0204 | 0.0120 | 0.0024 | 0.0046 | 0.0041 | 1.0058 |
| 240 | 21 | | 1,20 | 0.0174 | 0.5972 | 0.0024 | 0.1927 | 0.1664 | 0.0187 | 0.0079 | 0.0024 | 0.0017 | 0.0071 | 1.0139 |
| 241 | 22 | | 1,20 | 0.0281 | 0.5722 | 0.0024 | 0.1918 | 0.1862 | 0.0164 | 0.0057 | 0.0027 | 0.0020 | 0.0060 | 1.0134 |
| 242 | 23 | | 1,20 | 0.0329 | 0.5619 | 0.0000 | 0.1908 | 0.2004 | 0.0148 | 0.0057 | 0.0009 | 0.0020 | 0.0040 | 1.0135 |
| 243 | 100 | 12 | 1,3 | 0.0375 | 0.6884 | 0.0004 | 0.1601 | 0.0510 | 0.0297 | 0.0028 | 0.0176 | 0.0017 | 0.0030 | 0.9923 |
| 244 | 101 | 13 | 1,3 | 0.0321 | 0.6989 | 0.0009 | 0.1643 | 0.0431 | 0.0305 | 0.0025 | 0.0218 | 0.0012 | 0.0021 | 0.9973 |
| 245 | 105 | 17 | 1,3 | 0.0436 | 0.6845 | 0.0009 | 0.1576 | 0.0471 | 0.0311 | 0.0000 | 0.0187 | 0.0013 | 0.0000 | 0.9849 |
| 246 | 106 | 18 | 1,3 | 0.0522 | 0.6848 | 0.0008 | 0.1502 | 0.0495 | 0.0290 | 0.0039 | 0.0147 | 0.0018 | 0.0026 | 0.9896 |
| 247 | 75 | | 1,4 | 0.0631 | 0.6636 | 0.0017 | 0.1377 | 0.0420 | 0.0410 | 0.0033 | 0.0341 | 0.0022 | 0.0008 | 0.9895 |
| 248 | 76 | | 1,4 | 0.0616 | 0.6697 | 0.0009 | 0.1360 | 0.0440 | 0.0474 | 0.0033 | 0.0353 | 0.0017 | 0.0010 | 1.0009 |
| 249 | 77 | | 1,4 | 0.0605 | 0.6658 | 0.0009 | 0.1362 | 0.0431 | 0.0459 | 0.0029 | 0.0344 | 0.0018 | 0.0010 | 0.9926 |
| 250 | 78 | | 1,4 | 0.0588 | 0.6665 | 0.0010 | 0.1386 | 0.0419 | 0.0429 | 0.0038 | 0.0358 | 0.0021 | 0.0012 | 0.9927 |
| 251 | 79 | | 1,5 | 0.0121 | 0.5754 | 0.0000 | 0.1970 | 0.1807 | 0.0208 | 0.0041 | 0.0055 | 0.0025 | 0.0047 | 1.0029 |
| 252 | 80 | | 1,5 | 0.0090 | 0.5728 | 0.0000 | 0.1996 | 0.1758 | 0.0198 | 0.0037 | 0.0130 | 0.0029 | 0.0042 | 1.0009 |
| 253 | 81 | | 1,5 | 0.0000 | 0.5875 | 0.0017 | 0.1996 | 0.1538 | 0.0258 | 0.0035 | 0.0148 | 0.0032 | 0.0051 | 0.9950 |
| 254 | 82 | | 1,5 | 0.0141 | 0.5658 | 0.0000 | 0.1962 | 0.1914 | 0.0232 | 0.0039 | 0.0050 | 0.0022 | 0.0045 | 1.0064 |
| 255 | 84 | | 1,6 | 0.0159 | 0.7094 | 0.0007 | 0.1811 | 0.0572 | 0.0245 | 0.0030 | 0.0065 | 0.0019 | 0.0017 | 1.0020 |
| 256 | 85 | | 1,6 | 0.0220 | 0.6966 | 0.0004 | 0.1748 | 0.0567 | 0.0267 | 0.0015 | 0.0093 | 0.0023 | 0.0017 | 0.9918 |
| 257 | 86 | | 1,6 | 0.0106 | 0.6755 | 0.0000 | 0.1873 | 0.0746 | 0.0197 | 0.0036 | 0.0203 | 0.0022 | 0.0025 | 0.9963 |
| 258 | 87 | | 1,7 | 0.0172 | 0.6833 | 0.0006 | 0.1626 | 0.0622 | 0.0482 | 0.0031 | 0.0061 | 0.0016 | 0.0013 | 0.9862 |
| 259 | 88 | | 1,7 | 0.0278 | 0.6790 | 0.0002 | 0.1554 | 0.0639 | 0.0481 | 0.0032 | 0.0030 | 0.0024 | 0.0019 | 0.9849 |
| 260 | 89 | 1 | 1,7 | 0.0201 | 0.6854 | 0.0002 | 0.1609 | 0.0653 | 0.0491 | 0.0031 | 0.0027 | 0.0018 | 0.0016 | 0.9902 |
| 261 | 93 | 5 | 1,9 | 0.0310 | 0.6164 | 0.0003 | 0.1766 | 0.1324 | 0.0220 | 0.0051 | 0.0056 | 0.0031 | 0.0059 | 0.9985 |
| 262 | 94 | 6 | 1,9 | 0.0237 | 0.5975 | 0.0003 | 0.1745 | 0.1321 | 0.0223 | 0.0046 | 0.0088 | 0.0009 | 0.0058 | 0.9705 |
| 263 | 95 | 7 | 1,9 | 0.0308 | 0.5985 | 0.0001 | 0.1746 | 0.1411 | 0.0225 | 0.0047 | 0.0076 | 0.0014 | 0.0053 | 0.9866 |
| 264 | 27 | | 2,1 | 0.0418 | 0.7111 | 0.0003 | 0.1452 | 0.0369 | 0.0486 | 0.0039 | 0.0109 | 0.0022 | 0.0035 | 1.0044 |
| 265 | 28 | | 2,1 | 0.0359 | 0.7007 | 0.0003 | 0.1463 | 0.0382 | 0.0510 | 0.0042 | 0.0173 | 0.0016 | 0.0031 | 0.9986 |
| 266 | 29 | | 2,1 | 0.0302 | 0.7212 | 0.0000 | 0.1520 | 0.0440 | 0.0545 | 0.0037 | 0.0049 | 0.0015 | 0.0027 | 1.0149 |
| 267 | 51 | | 2,10 | 0.0097 | 0.7250 | 0.0003 | 0.1920 | 0.0468 | 0.0141 | 0.0026 | 0.0190 | 0.0020 | 0.0035 | 1.0151 |
| 268 | 52 | | 2,10 | 0.0045 | 0.7315 | 0.0001 | 0.1949 | 0.0474 | 0.0170 | 0.0025 | 0.0183 | 0.0015 | 0.0027 | 1.0204 |
| 269 | 53 | | 2,10 | 0.0066 | 0.7304 | 0.0000 | 0.1924 | 0.0454 | 0.0124 | 0.0028 | 0.0105 | 0.0016 | 0.0032 | 1.0053 |
| 270 | 54 | | 2,11 | 0.0000 | 0.7402 | 0.0004 | 0.1941 | 0.0419 | 0.0221 | 0.0046 | 0.0050 | 0.0020 | 0.0031 | 1.0133 |
| 271 | 55 | | 2,11 | 0.0000 | 0.7391 | 0.0013 | 0.1940 | 0.0377 | 0.0247 | 0.0039 | 0.0067 | 0.0018 | 0.0017 | 1.0109 |
| 272 | 56 | | 2,12 | 0.0228 | 0.7211 | 0.0010 | 0.1823 | 0.0405 | 0.0152 | 0.0025 | 0.0170 | 0.0023 | 0.0020 | 1.0067 |
| 273 | 57 | | 2,12 | 0.0192 | 0.7310 | 0.0008 | 0.1869 | 0.0384 | 0.0161 | 0.0000 | 0.0154 | 0.0020 | 0.0008 | 1.0106 |
| 274 | 58 | | 2,12 | 0.0036 | 0.7401 | 0.0007 | 0.1939 | 0.0372 | 0.0173 | 0.0029 | 0.0121 | 0.0023 | 0.0017 | 1.0118 |
| 275 | 59 | | 2,13 | 0.0292 | 0.6972 | 0.0007 | 0.1582 | 0.0476 | 0.0328 | 0.0047 | 0.0136 | 0.0024 | 0.0095 | 0.9958 |
| 276 | 60 | | 2,13 | 0.0239 | 0.6866 | 0.0005 | 0.1634 | 0.0570 | 0.0321 | 0.0048 | 0.0139 | 0.0025 | 0.0082 | 0.9930 |
| 277 | 61 | | 2,13 | 0.0311 | 0.6836 | 0.0004 | 0.1611 | 0.0602 | 0.0322 | 0.0060 | 0.0187 | 0.0025 | 0.0070 | 1.0028 |
| 278 | 62 | | 2,13 | 0.0313 | 0.6835 | 0.0009 | 0.1619 | 0.0602 | 0.0326 | 0.0041 | 0.0157 | 0.0025 | 0.0066 | 0.9993 |
| 279 | 63 | | 2,14 | 0.0311 | 0.6262 | 0.0002 | 0.1763 | 0.1371 | 0.0311 | 0.0043 | 0.0017 | 0.0022 | 0.0032 | 1.0135 |
| 280 | 64 | | 2,14 | 0.0332 | 0.6297 | 0.0000 | 0.1744 | 0.1340 | 0.0324 | 0.0030 | 0.0009 | 0.0028 | 0.0039 | 1.0142 |

| | | | | | | | | | | | | | |
|---|---|---|---|---|---|---|---|---|---|---|---|---|---|
| 281 | 65 | 2,14 | 0.0432 | 0.6178 | 0.0000 | 0.1736 | 0.1341 | 0.0140 | 0.0044 | 0.0000 | 0.0013 | 0.0034 | 0.9919 |
| 282 | 67 | 2,15 | 0.0000 | 0.6720 | 0.0010 | 0.2114 | 0.1020 | 0.0117 | 0.0050 | 0.0036 | 0.0041 | 0.0052 | 1.0160 |
| 283 | 68 | 2,16 | 0.0093 | 0.7199 | 0.0005 | 0.1791 | 0.0339 | 0.0290 | 0.0033 | 0.0250 | 0.0011 | 0.0025 | 1.0035 |
| 284 | 69 | 2,16 | 0.0109 | 0.7193 | 0.0000 | 0.1813 | 0.0388 | 0.0285 | 0.0020 | 0.0255 | 0.0021 | 0.0026 | 1.0110 |
| 285 | 70 | 2,16 | 0.0029 | 0.7134 | 0.0003 | 0.1828 | 0.0408 | 0.0312 | 0.0035 | 0.0261 | 0.0015 | 0.0025 | 1.0049 |
| 286 | 71 | 2,17 | 0.0000 | 0.7403 | 0.0000 | 0.1910 | 0.0285 | 0.0160 | 0.0033 | 0.0159 | 0.0011 | 0.0043 | 1.0004 |
| 287 | 72 | 2,17 | 0.0005 | 0.7406 | 0.0004 | 0.1924 | 0.0277 | 0.0132 | 0.0023 | 0.0205 | 0.0004 | 0.0045 | 1.0025 |
| 288 | 73 | 2,17 | 0.0000 | 0.7458 | 0.0001 | 0.1916 | 0.0283 | 0.0170 | 0.0025 | 0.0156 | 0.0010 | 0.0045 | 1.0062 |
| 289 | 74 | 2,18 | 0.0097 | 0.6976 | 0.0006 | 0.1728 | 0.0426 | 0.0365 | 0.0038 | 0.0221 | 0.0017 | 0.0017 | 0.9891 |
| 290 | 75 | 2,18 | 0.0371 | 0.6815 | 0.0003 | 0.1603 | 0.0593 | 0.0285 | 0.0032 | 0.0109 | 0.0017 | 0.0030 | 0.9857 |
| 291 | 76 | 2,18 | 0.0549 | 0.6972 | 0.0000 | 0.1532 | 0.0608 | 0.0273 | 0.0050 | 0.0024 | 0.0023 | 0.0029 | 1.0064 |
| 292 | 30 | 2,2 | 0.0227 | 0.6111 | 0.0004 | 0.1883 | 0.1605 | 0.0225 | 0.0055 | 0.0006 | 0.0017 | 0.0053 | 1.0187 |
| 293 | 31 | 2,2 | 0.0286 | 0.6117 | 0.0003 | 0.1819 | 0.1506 | 0.0233 | 0.0055 | 0.0002 | 0.0020 | 0.0040 | 1.0081 |
| 294 | 32 | 2,2 | 0.0263 | 0.6498 | 0.0000 | 0.1756 | 0.1156 | 0.0318 | 0.0042 | 0.0015 | 0.0020 | 0.0032 | 1.0102 |
| 295 | 33 | 2,3 | 0.0042 | 0.7297 | 0.0005 | 0.1957 | 0.0502 | 0.0125 | 0.0044 | 0.0088 | 0.0019 | 0.0024 | 1.0104 |
| 296 | 34 | 2,3 | 0.0065 | 0.7103 | 0.0040 | 0.1975 | 0.0601 | 0.0215 | 0.0038 | 0.0187 | 0.0030 | 0.0021 | 1.0275 |
| 297 | 35 | 2,3 | 0.0183 | 0.7300 | 0.0000 | 0.1907 | 0.0409 | 0.0000 | 0.0034 | 0.0094 | 0.0019 | 0.0028 | 0.9975 |
| 298 | 36 | 2,4 | 0.0013 | 0.7507 | 0.0007 | 0.1903 | 0.0308 | 0.0224 | 0.0047 | 0.0108 | 0.0014 | 0.0016 | 1.0146 |
| 299 | 38 | 2,4 | 0.0029 | 0.7454 | 0.0006 | 0.1880 | 0.0365 | 0.0246 | 0.0029 | 0.0050 | 0.0015 | 0.0022 | 1.0096 |
| 300 | 39 | 2,5 | 0.0553 | 0.6860 | 0.0007 | 0.1496 | 0.0674 | 0.0355 | 0.0049 | 0.0012 | 0.0024 | 0.0012 | 1.0041 |
| 301 | 40 | 2,5 | 0.0600 | 0.6863 | 0.0009 | 0.1467 | 0.0620 | 0.0352 | 0.0038 | 0.0016 | 0.0029 | 0.0012 | 1.0005 |
| 302 | 41 | 2,5 | 0.0549 | 0.6881 | 0.0007 | 0.1480 | 0.0658 | 0.0417 | 0.0032 | 0.0013 | 0.0029 | 0.0011 | 1.0077 |
| 303 | 42 | 2,6 | 0.0208 | 0.6921 | 0.0005 | 0.1786 | 0.0795 | 0.0307 | 0.0033 | 0.0043 | 0.0029 | 0.0030 | 1.0158 |
| 304 | 43 | 2,6 | 0.0245 | 0.7063 | 0.0007 | 0.1745 | 0.0409 | 0.0280 | 0.0040 | 0.0380 | 0.0019 | 0.0025 | 1.0212 |
| 305 | 44 | 2,6 | 0.0100 | 0.6792 | 0.0001 | 0.1825 | 0.0869 | 0.0355 | 0.0035 | 0.0080 | 0.0029 | 0.0024 | 1.0110 |
| 306 | 45 | 2,7 | 0.0052 | 0.7317 | 0.0002 | 0.1916 | 0.0450 | 0.0174 | 0.0022 | 0.0140 | 0.0013 | 0.0037 | 1.0122 |
| 307 | 46 | 2,7 | 0.0066 | 0.7380 | 0.0007 | 0.1909 | 0.0421 | 0.0174 | 0.0024 | 0.0115 | 0.0014 | 0.0043 | 1.0153 |
| 308 | 47 | 2,7 | 0.0093 | 0.7228 | 0.0006 | 0.1881 | 0.0442 | 0.0167 | 0.0026 | 0.0161 | 0.0015 | 0.0039 | 1.0058 |
| 309 | 48 | 2,8 | 0.0280 | 0.6042 | 0.0000 | 0.1824 | 0.1484 | 0.0218 | 0.0060 | 0.0074 | 0.0023 | 0.0030 | 1.0035 |
| 310 | 49 | 2,8 | 0.0281 | 0.6087 | 0.0000 | 0.1812 | 0.1469 | 0.0225 | 0.0056 | 0.0048 | 0.0018 | 0.0035 | 1.0030 |
| 311 | 50 | 2,8 | 0.0201 | 0.6175 | 0.0001 | 0.1876 | 0.1443 | 0.0209 | 0.0073 | 0.0042 | 0.0021 | 0.0029 | 1.0071 |
| 312 | 51 | 3,10 | 0.0207 | 0.6974 | 0.0006 | 0.1699 | 0.0420 | 0.0243 | 0.0045 | 0.0177 | 0.0017 | 0.0037 | 0.9824 |
| 313 | 52 | 3,10 | 0.0202 | 0.7106 | 0.0002 | 0.1716 | 0.0412 | 0.0270 | 0.0039 | 0.0184 | 0.0010 | 0.0031 | 0.9972 |
| 314 | 53 | 3,10 | 0.0367 | 0.6970 | 0.0002 | 0.1638 | 0.0497 | 0.0204 | 0.0043 | 0.0165 | 0.0015 | 0.0055 | 0.9957 |
| 315 | 54 | 3,11 | 0.0449 | 0.6617 | 0.0008 | 0.1512 | 0.0634 | 0.0385 | 0.0021 | 0.0128 | 0.0030 | 0.0016 | 0.9800 |
| 316 | 55 | 3,11 | 0.0266 | 0.6735 | 0.0010 | 0.1632 | 0.0604 | 0.0396 | 0.0030 | 0.0139 | 0.0028 | 0.0008 | 0.9849 |
| 317 | 56 | 3,11 | 0.0284 | 0.6753 | 0.0015 | 0.1604 | 0.0555 | 0.0417 | 0.0034 | 0.0168 | 0.0026 | 0.0011 | 0.9867 |
| 318 | 57 | 3,12 | 0.0246 | 0.6803 | 0.0000 | 0.1733 | 0.0782 | 0.0283 | 0.0033 | 0.0014 | 0.0024 | 0.0017 | 0.9935 |
| 319 | 58 | 3,12 | 0.0266 | 0.6714 | 0.0025 | 0.1757 | 0.0805 | 0.0293 | 0.0034 | 0.0089 | 0.0024 | 0.0016 | 1.0023 |
| 320 | 59 | 3,12 | 0.0278 | 0.6719 | 0.0007 | 0.1731 | 0.0793 | 0.0255 | 0.0027 | 0.0044 | 0.0019 | 0.0017 | 0.9890 |
| 321 | 60 | 3,12 | 0.0319 | 0.6704 | 0.0006 | 0.1691 | 0.0738 | 0.0244 | 0.0029 | 0.0060 | 0.0022 | 0.0016 | 0.9828 |
| 322 | 61 | 3,13 | 0.0233 | 0.6772 | 0.0005 | 0.1721 | 0.0781 | 0.0303 | 0.0037 | 0.0042 | 0.0017 | 0.0024 | 0.9935 |
| 323 | 62 | 3,13 | 0.0158 | 0.6833 | 0.0002 | 0.1783 | 0.0759 | 0.0248 | 0.0049 | 0.0012 | 0.0019 | 0.0024 | 0.9887 |
| 324 | 63 | 3,13 | 0.0245 | 0.6884 | 0.0000 | 0.1725 | 0.0725 | 0.0257 | 0.0054 | 0.0025 | 0.0018 | 0.0025 | 0.9958 |
| 325 | 64 | 3,14,A | 0.0443 | 0.6848 | 0.0009 | 0.1469 | 0.0404 | 0.0395 | 0.0030 | 0.0171 | 0.0020 | 0.0025 | 0.9813 |
| 326 | 66 | 3,14,A | 0.0427 | 0.6888 | 0.0009 | 0.1460 | 0.0423 | 0.0433 | 0.0032 | 0.0124 | 0.0014 | 0.0019 | 0.9830 |
| 327 | 68 | 3,14,B | 0.0355 | 0.6862 | 0.0032 | 0.1498 | 0.0430 | 0.0487 | 0.0030 | 0.0170 | 0.0012 | 0.0036 | 0.9911 |
| 328 | 70 | 3,14,C | 0.0345 | 0.6972 | 0.0007 | 0.1510 | 0.0375 | 0.0405 | 0.0035 | 0.0130 | 0.0014 | 0.0034 | 0.9828 |
| 329 | 72 | 3,15 | 0.0035 | 0.7109 | 0.0004 | 0.1921 | 0.0467 | 0.0145 | 0.0000 | 0.0114 | 0.0016 | 0.0015 | 0.9826 |
| 330 | 73 | 3,15 | 0.0000 | 0.7145 | 0.0000 | 0.1931 | 0.0457 | 0.0129 | 0.0035 | 0.0125 | 0.0020 | 0.0035 | 0.9877 |
| 331 | 74 | 3,15 | 0.0100 | 0.7084 | 0.0004 | 0.1854 | 0.0464 | 0.0144 | 0.0030 | 0.0150 | 0.0014 | 0.0035 | 0.9879 |
| 332 | 16 | 3,2 | 0.0040 | 0.7286 | 0.0019 | 0.1873 | 0.0281 | 0.0207 | 0.0024 | 0.0147 | 0.0017 | 0.0016 | 0.9911 |
| 333 | 22 | 3,2 | 0.0202 | 0.7273 | 0.0015 | 0.1778 | 0.0290 | 0.0213 | 0.0029 | 0.0168 | 0.0020 | 0.0016 | 1.0004 |
| 334 | 23 | 3,2 | 0.0226 | 0.7317 | 0.0011 | 0.1780 | 0.0287 | 0.0184 | 0.0028 | 0.0190 | 0.0009 | 0.0017 | 1.0049 |
| 335 | 24 | 3,2 | 0.0213 | 0.7260 | 0.0011 | 0.1738 | 0.0305 | 0.0215 | 0.0067 | 0.0172 | 0.0015 | 0.0023 | 1.0020 |
| 336 | 26 | 3,3 | 0.0272 | 0.6473 | 0.0008 | 0.1684 | 0.0591 | 0.0293 | 0.0033 | 0.0440 | 0.0029 | 0.0017 | 0.9840 |
| 337 | 28 | 3,3 | 0.0281 | 0.6737 | 0.0000 | 0.1642 | 0.0588 | 0.0288 | 0.0083 | 0.0287 | 0.0023 | 0.0042 | 0.9971 |
| 338 | 29 | 3,4 | 0.0048 | 0.7332 | 0.0007 | 0.1930 | 0.0396 | 0.0117 | 0.0029 | 0.0077 | 0.0017 | 0.0025 | 0.9978 |
| 339 | 30 | 3,4 | 0.0053 | 0.7339 | 0.0005 | 0.1946 | 0.0427 | 0.0113 | 0.0029 | 0.0085 | 0.0020 | 0.0024 | 1.0040 |
| 340 | 32 | 3,4 | 0.0120 | 0.7298 | 0.0010 | 0.1899 | 0.0399 | 0.0112 | 0.0035 | 0.0118 | 0.0021 | 0.0027 | 1.0040 |
| 341 | 33 | 3,4 | 0.0192 | 0.7250 | 0.0001 | 0.1838 | 0.0396 | 0.0117 | 0.0019 | 0.0106 | 0.0018 | 0.0021 | 0.9958 |
| 342 | 34 | 3,5 | 0.0000 | 0.6536 | 0.0000 | 0.1941 | 0.1178 | 0.0145 | 0.0081 | 0.0050 | 0.0002 | 0.0141 | 1.0072 |
| 343 | 35 | 3,5 | 0.0075 | 0.6739 | 0.0002 | 0.1856 | 0.0894 | 0.0142 | 0.0049 | 0.0147 | 0.0016 | 0.0153 | 1.0072 |
| 344 | 36 | 3,5 | 0.0000 | 0.6820 | 0.0000 | 0.1901 | 0.0906 | 0.0123 | 0.0051 | 0.0080 | 0.0008 | 0.0163 | 1.0052 |
| 345 | 37 | 3,6 | 0.0161 | 0.5996 | 0.0000 | 0.1971 | 0.1652 | 0.0054 | 0.0046 | 0.0011 | 0.0015 | 0.0096 | 1.0002 |
| 346 | 38 | 3,6 | 0.0048 | 0.6216 | 0.0000 | 0.2034 | 0.1561 | 0.0103 | 0.0043 | 0.0021 | 0.0016 | 0.0093 | 1.0136 |
| 347 | 41 | 3,7 | 0.0576 | 0.6850 | 0.0016 | 0.1535 | 0.0567 | 0.0254 | 0.0028 | 0.0065 | 0.0018 | 0.0000 | 0.9910 |
| 348 | 42 | 3,7 | 0.0616 | 0.6771 | 0.0013 | 0.1508 | 0.0602 | 0.0268 | 0.0026 | 0.0099 | 0.0030 | 0.0011 | 0.9942 |
| 349 | 43 | 3,7 | 0.0501 | 0.6893 | 0.0010 | 0.1566 | 0.0484 | 0.0245 | 0.0040 | 0.0080 | 0.0042 | 0.0010 | 0.9871 |
| 350 | 44 | 3,8 | 0.0238 | 0.6806 | 0.0010 | 0.1618 | 0.0632 | 0.0386 | 0.0032 | 0.0077 | 0.0021 | 0.0063 | 0.9882 |
| 351 | 45 | 3,8 | 0.0359 | 0.6689 | 0.0021 | 0.1550 | 0.0664 | 0.0371 | 0.0027 | 0.0049 | 0.0027 | 0.0074 | 0.9831 |
| 352 | 46 | 3,8 | 0.0333 | 0.6519 | 0.0058 | 0.1599 | 0.0753 | 0.0340 | 0.0036 | 0.0075 | 0.0018 | 0.0097 | 0.9827 |
| 353 | 49 | 3,9 | 0.0389 | 0.6671 | 0.0012 | 0.1604 | 0.0621 | 0.0281 | 0.0033 | 0.0140 | 0.0024 | 0.0022 | 0.9797 |
| 354 | 17 1x94 | 2#14 | 0.0112 | 0.7043 | 0.0006 | 0.1743 | 0.0518 | 0.0338 | 0.0033 | 0.0059 | 0.0020 | 0.0027 | 0.9898 |
| 355 | 18 1x94 | 2#14 | 0.0050 | 0.7314 | 0.0036 | 0.1753 | 0.0252 | 0.0369 | 0.0030 | 0.0077 | 0.0009 | 0.0036 | 0.9926 |
| 356 | 19 1x94 | 2#14 | 0.0087 | 0.7210 | 0.0014 | 0.1757 | 0.0384 | 0.0331 | 0.0040 | 0.0090 | 0.0014 | 0.0031 | 0.9958 |
| 357 | 20 1x94 | 2#14 | 0.0000 | 0.7408 | 0.0000 | 0.2007 | 0.0399 | 0.0120 | 0.0033 | 0.0119 | 0.0008 | 0.0100 | 1.0195 |
| 358 | 23 1x94 | 2#2 | 0.0307 | 0.6482 | 0.0326 | 0.1768 | 0.0604 | 0.0172 | 0.0032 | 0.0048 | 0.0027 | 0.0419 | 1.0184 |
| 359 | 24 1x94 | 2#2 | 0.0179 | 0.7066 | 0.0002 | 0.1840 | 0.0627 | 0.0180 | 0.0020 | 0.0045 | 0.0027 | 0.0033 | 1.0021 |
| 360 | 25 1x94 | 2#2 | 0.0235 | 0.6958 | 0.0052 | 0.1798 | 0.0649 | 0.0167 | 0.0030 | 0.0051 | 0.0021 | 0.0111 | 1.0071 |
| 361 | 26 1x94 | 2#5 | 0.0056 | 0.5497 | 0.0000 | 0.2067 | 0.2116 | 0.0198 | 0.0043 | 0.0022 | 0.0040 | 0.0057 | 1.0096 |
| 362 | 27 1x94 | 2#5 | 0.0056 | 0.5617 | 0.0000 | 0.2066 | 0.2007 | 0.0180 | 0.0029 | 0.0019 | 0.0037 | 0.0059 | 1.0069 |
| 363 | 28 1x94 | 2#5 | 0.0001 | 0.5723 | 0.0003 | 0.2125 | 0.1993 | 0.0173 | 0.0035 | 0.0021 | 0.0041 | 0.0063 | 1.0178 |
| 364 | 29 1x94 | 2#10 | 0.0201 | 0.6147 | 0.0003 | 0.1919 | 0.1505 | 0.0187 | 0.0049 | 0.0035 | 0.0028 | 0.0040 | 1.0114 |
| 365 | 30 1x94 | 2#10 | 0.0246 | 0.6218 | 0.0001 | 0.1859 | 0.1414 | 0.0218 | 0.0041 | 0.0051 | 0.0023 | 0.0040 | 1.0111 |
| 366 | 31 1x94 | 2#10 | 0.0301 | 0.6168 | 0.0004 | 0.1791 | 0.1357 | 0.0227 | 0.0032 | 0.0035 | 0.0024 | 0.0045 | 0.9985 |
| 367 | 33 1x94 | 2#11 | 0.0148 | 0.6392 | 0.0000 | 0.1865 | 0.1236 | 0.0291 | 0.0026 | 0.0082 | 0.0022 | 0.0037 | 1.0099 |
| 368 | 34 1x94 | 2#11 | 0.0247 | 0.6185 | 0.0009 | 0.1700 | 0.0754 | 0.0321 | 0.0033 | 0.0602 | 0.0019 | 0.0027 | 0.9897 |
| 369 | 35 1x94 | 2#11 | 0.0262 | 0.6256 | 0.0010 | 0.1676 | 0.0769 | 0.0354 | 0.0025 | 0.0518 | 0.0021 | 0.0038 | 0.9929 |
| 370 | 36 1x94 | 2#12 | 0.0000 | 0.5284 | 0.0002 | 0.2161 | 0.2349 | 0.0175 | 0.0046 | 0.0026 | 0.0029 | 0.0070 | 1.0142 |
| 371 | 37 1x94 | 2#12 | 0.0006 | 0.5288 | 0.0002 | 0.2119 | 0.2301 | 0.0180 | 0.0039 | 0.0021 | 0.0031 | 0.0060 | 1.0046 |
| 372 | 42 1x94 | 2#18 | 0.0053 | 0.7293 | 0.0000 | 0.1880 | 0.0539 | 0.0097 | 0.0039 | 0.0048 | 0.0015 | 0.0134 | 1.0099 |
| 373 | 43 1x94 | 2#18 | 0.0074 | 0.7279 | 0.0002 | 0.1882 | 0.0556 | 0.0090 | 0.0036 | 0.0044 | 0.0014 | 0.0129 | 1.0105 |
| 374 | 44 1x94 | 2#18 | 0.0000 | 0.7314 | 0.0000 | 0.1918 | 0.0515 | 0.0086 | 0.0048 | 0.0070 | 0.0010 | 0.0127 | 1.0087 |
| 375 | 45 1x94 | 2#20 | 0.0309 | 0.7414 | 0.0000 | 0.1597 | 0.0349 | 0.0110 | 0.0044 | 0.0031 | 0.0006 | 0.0236 | 1.0095 |
| 376 | 46 1x94 | 2#20 | 0.0429 | 0.7203 | 0.0000 | 0.1567 | 0.0446 | 0.0101 | 0.0062 | 0.0010 | 0.0008 | 0.0150 | 0.9975 |

| | | | | | | | | | | | | | | |
|---|---|---|---|---|---|---|---|---|---|---|---|---|---|---|
| 377 | 48 | 1x94 | 2#21 | 0.0037 | 0.6053 | 0.0000 | 0.2068 | 0.1622 | 0.0107 | 0.0045 | 0.0091 | 0.0024 | 0.0065 | 1.0113 |
| 378 | 49 | 1x94 | 2#21 | 0.0081 | 0.5988 | 0.0001 | 0.2010 | 0.1550 | 0.0089 | 0.0035 | 0.0068 | 0.0027 | 0.0067 | 0.9917 |
| 379 | 51 | 1x94 | 2#33 | 0.0000 | 0.6717 | 0.0000 | 0.2000 | 0.1033 | 0.0087 | 0.0047 | 0.0093 | 0.0021 | 0.0136 | 1.0133 |
| 380 | 52 | 1x94 | 2#33 | -0.0003 | 0.6782 | 0.0000 | 0.1915 | 0.0900 | 0.0074 | 0.0052 | 0.0066 | 0.0026 | 0.0179 | 0.9991 |
| 381 | 53 | 1x94 | 2#35 | 0.0366 | 0.6131 | 0.0000 | 0.1751 | 0.1402 | 0.0177 | 0.0054 | 0.0061 | 0.0014 | 0.0069 | 1.0024 |
| 382 | 54 | 1x94 | 2#35 | 0.0338 | 0.6193 | 0.0001 | 0.1827 | 0.1482 | 0.0168 | 0.0045 | 0.0020 | 0.0019 | 0.0059 | 1.0153 |
| 383 | 55 | 1x94 | 2#39 | 0.0022 | 0.7287 | 0.0000 | 0.1890 | 0.0604 | 0.0081 | 0.0044 | 0.0039 | 0.0015 | 0.0185 | 1.0167 |
| 384 | 56 | 1x94 | 2#39 | 0.0067 | 0.7219 | 0.0000 | 0.1861 | 0.0625 | 0.0076 | 0.0047 | 0.0048 | 0.0013 | 0.0180 | 1.0136 |
| 385 | 57 | 1x94 | 2#41 | 0.0174 | 0.6862 | 0.0000 | 0.1875 | 0.0921 | 0.0161 | 0.0040 | 0.0053 | 0.0025 | 0.0076 | 1.0187 |
| 386 | 58 | 1x94 | 2#41 | 0.0177 | 0.6729 | 0.0003 | 0.1863 | 0.0959 | 0.0138 | 0.0041 | 0.0021 | 0.0027 | 0.0074 | 1.0031 |
| 387 | 59 | 1x94 | 2#41 | 0.0585 | 0.6662 | 0.0001 | 0.1627 | 0.0957 | 0.0129 | 0.0050 | 0.0033 | 0.0017 | 0.0048 | 1.0110 |
| 388 | 60 | 1x94 | 3#1 | 0.0161 | 0.5700 | 0.0001 | 0.1936 | 0.1838 | 0.0165 | 0.0051 | 0.0014 | 0.0019 | 0.0065 | 0.9950 |
| 389 | 61 | 1x94 | 3#1 | 0.0207 | 0.5929 | 0.0014 | 0.1901 | 0.1622 | 0.0157 | 0.0053 | 0.0022 | 0.0028 | 0.0081 | 1.0014 |
| 390 | 62 | 1x94 | 3#2 | 0.0236 | 0.6866 | 0.0000 | 0.1768 | 0.0731 | 0.0131 | 0.0043 | 0.0089 | 0.0013 | 0.0089 | 0.9966 |
| 391 | 63 | 1x94 | 3#2 | -0.0005 | 0.6850 | 0.0000 | 0.1969 | 0.0914 | 0.0128 | 0.0046 | 0.0049 | 0.0025 | 0.0094 | 1.0070 |
| 392 | 66 | 1x94 | 3#3 | 0.0077 | 0.6717 | 0.0000 | 0.1893 | 0.0912 | 0.0200 | 0.0052 | 0.0016 | 0.0032 | 0.0028 | 0.9927 |
| 393 | 67 | 1x94 | 3#3 | 0.0111 | 0.6636 | 0.0001 | 0.1876 | 0.0980 | 0.0208 | 0.0054 | 0.0013 | 0.0036 | 0.0030 | 0.9946 |
| 394 | 68 | 1x94 | 3#8 | 0.0073 | 0.6219 | 0.0000 | 0.1989 | 0.1483 | 0.0124 | 0.0056 | 0.0016 | 0.0021 | 0.0074 | 1.0055 |
| 395 | 69 | 1x94 | 3#8 | 0.0044 | 0.6254 | 0.0000 | 0.2010 | 0.1487 | 0.0128 | 0.0053 | 0.0011 | 0.0023 | 0.0081 | 1.0091 |
| 396 | 70 | 1x94 | 3#16 | 0.0289 | 0.6888 | 0.0002 | 0.1799 | 0.0765 | 0.0091 | 0.0042 | 0.0042 | 0.0022 | 0.0061 | 1.0001 |
| 397 | 71 | 1x94 | 3#16 | 0.0218 | 0.7106 | 0.0004 | 0.1847 | 0.0650 | 0.0108 | 0.0036 | 0.0065 | 0.0020 | 0.0066 | 1.0120 |
| 398 | 72 | 1x94 | 3#16 | 0.0000 | 0.7296 | 0.0001 | 0.2024 | 0.0581 | 0.0106 | 0.0040 | 0.0074 | 0.0012 | 0.0058 | 1.0192 |
| 399 | 73 | 1x94 | 3#18 | 0.0036 | 0.7064 | 0.0004 | 0.1894 | 0.0699 | 0.0166 | 0.0038 | 0.0061 | 0.0019 | 0.0104 | 1.0085 |
| 400 | 74 | 1x94 | 3#18 | 0.0067 | 0.6950 | 0.0000 | 0.1869 | 0.0768 | 0.0165 | 0.0037 | 0.0053 | 0.0021 | 0.0107 | 1.0036 |
| 401 | 75 | 1x94 | 3#18 | 0.0000 | 0.6887 | 0.0000 | 0.1909 | 0.0699 | 0.0156 | 0.0041 | 0.0164 | 0.0023 | 0.0108 | 0.9986 |
| 402 | 76 | 1x94 | 3#19 | 0.0291 | 0.6138 | 0.0000 | 0.1805 | 0.1338 | 0.0064 | 0.0071 | 0.0009 | 0.0013 | 0.0071 | 0.9800 |
| 403 | 77 | 1x94 | 3#19 | 0.0329 | 0.6201 | 0.0000 | 0.1849 | 0.1457 | 0.0067 | 0.0069 | 0.0009 | 0.0013 | 0.0067 | 1.0061 |
| 404 | 9 | 2x94 | JB-1#2 | 0.0000 | 0.7063 | 0.0008 | 0.2009 | 0.0744 | 0.0135 | 0.0038 | 0.0059 | 0.0021 | 0.0040 | 1.0116 |
| 405 | 10 | 2x94 | JB-1#2 | 0.0000 | 0.6965 | 0.0003 | 0.2007 | 0.0821 | 0.0149 | 0.0039 | 0.0046 | 0.0028 | 0.0033 | 1.0090 |
| 406 | 11 | 2x94 | JB-1#2 | 0.0069 | 0.6945 | 0.0005 | 0.1963 | 0.0831 | 0.0146 | 0.0031 | 0.0032 | 0.0026 | 0.0035 | 1.0083 |
| 407 | 12 | 2x94 | JB-1#2 | 0.0000 | 0.6510 | 0.0004 | 0.2097 | 0.1134 | 0.0156 | 0.0033 | 0.0059 | 0.0050 | 0.0025 | 1.0068 |
| 408 | 29 | 2x94 | 3#25 | 0.0257 | 0.6723 | 0.0002 | 0.1878 | 0.1013 | 0.0082 | 0.0052 | 0.0025 | 0.0034 | 0.0065 | 1.0131 |
| 409 | 30 | 2x94 | 3#25 | 0.0277 | 0.6442 | 0.0000 | 0.1851 | 0.1194 | 0.0074 | 0.0054 | 0.0018 | 0.0027 | 0.0074 | 1.0011 |
| 410 | 31 | 2x94 | 3#25 | 0.0165 | 0.6636 | 0.0000 | 0.1890 | 0.0995 | 0.0097 | 0.0054 | 0.0040 | 0.0030 | 0.0063 | 0.9970 |
| 411 | 32 | 2x94 | 3#32 | 0.0134 | 0.5877 | 0.0000 | 0.1898 | 0.1616 | 0.0167 | 0.0059 | 0.0009 | 0.0016 | 0.0060 | 0.9836 |
| 412 | 33 | 2x94 | 3#32 | 0.0350 | 0.6313 | 0.0000 | 0.1757 | 0.1309 | 0.0226 | 0.0043 | 0.0017 | 0.0021 | 0.0051 | 1.0087 |
| 413 | 39 | 2x94 | 3#40 | 0.0000 | 0.6883 | 0.0000 | 0.1896 | 0.0664 | 0.0094 | 0.0048 | 0.0027 | 0.0022 | 0.0148 | 0.9782 |
| 414 | 43 | 2x94 | 3#51 | 0.0000 | 0.6506 | 0.0000 | 0.1980 | 0.1255 | 0.0141 | 0.0045 | 0.0074 | 0.0026 | 0.0098 | 1.0126 |
| 415 | DSPP 803D | | | | | | | | | | | | | |
| 416 | 31 | 9ix93 | 803DJB93-3-2 | 0.0026 | 0.6096 | 0.0000 | 0.1965 | 0.1610 | 0.0137 | 0.0049 | 0.0069 | 0.0024 | 0.0088 | 1.0062 |
| 417 | 32 | 9ix93 | 803DJB93-3-2 | 0.0000 | 0.5984 | 0.0001 | 0.2071 | 0.1695 | 0.0156 | 0.0051 | 0.0036 | 0.0022 | 0.0086 | 1.0102 |
| 418 | 33 | 9ix93 | 803DJB93-3-2 | 0.0000 | 0.6066 | 0.0001 | 0.2166 | 0.1698 | 0.0138 | 0.0057 | 0.0069 | 0.0030 | 0.0084 | 1.0308 |
| 419 | 34 | 9ix93 | 803DJB93-3-3 | 0.0044 | 0.5751 | 0.0000 | 0.1998 | 0.2017 | 0.0079 | 0.0038 | 0.0011 | 0.0015 | 0.0055 | 1.0008 |
| 420 | 35 | 9ix93 | 803DJB93-3-3 | 0.0034 | 0.5916 | 0.0000 | 0.2017 | 0.1869 | 0.0080 | 0.0040 | 0.0015 | 0.0015 | 0.0060 | 1.0047 |
| 421 | 36 | 9ix93 | 803DJB93-3-12 | 0.0012 | 0.7225 | 0.0000 | 0.1864 | 0.0598 | 0.0069 | 0.0055 | 0.0064 | 0.0017 | 0.0180 | 1.0086 |
| 422 | 39 | 9ix93 | 803DJB93-3-13 | 0.0000 | 0.6184 | 0.0003 | 0.2125 | 0.1458 | 0.0099 | 0.0039 | 0.0095 | 0.0024 | 0.0087 | 1.0113 |
| 423 | 40 | 9ix93 | 803DJB93-3-13 | 0.0000 | 0.6170 | 0.0008 | 0.2062 | 0.1433 | 0.0114 | 0.0031 | 0.0072 | 0.0032 | 0.0082 | 1.0005 |
| 424 | 87 | 9ix93 | 803DJB93-1-1 | 0.0235 | 0.7220 | 0.0002 | 0.1249 | 0.0415 | 0.0439 | 0.0025 | 0.0186 | 0.0019 | 0.0016 | 0.9806 |

| | | | | | | | | | | | | | |
|---|---|---|---|---|---|---|---|---|---|---|---|---|---|
| 425 | 88 | 9ix93 | 803DJB93-1-1 | 0.0198 | 0.7239 | 0.0001 | 0.1328 | 0.0430 | 0.0446 | 0.0024 | 0.0188 | 0.0015 | 0.0017 | 0.9888 |
| 426 | 89 | 9ix93 | 803DJB93-1-1 | 0.0193 | 0.7325 | 0.0013 | 0.1376 | 0.0442 | 0.0437 | 0.0028 | 0.0154 | 0.0015 | 0.0014 | 0.9996 |
| 427 | 90 | 9ix93 | 803DJB93-1-3 | 0.0106 | 0.6840 | 0.0000 | 0.1679 | 0.1017 | 0.0232 | 0.0058 | 0.0061 | 0.0017 | 0.0040 | 1.0050 |
| 428 | 91 | 9ix93 | 803DJB93-1-3 | 0.0046 | 0.6624 | 0.0007 | 0.1685 | 0.0999 | 0.0451 | 0.0040 | 0.0028 | 0.0030 | 0.0040 | 0.9950 |
| 429 | 92 | 9ix93 | 803DJB93-1-5 | 0.0000 | 0.7227 | 0.0015 | 0.2094 | 0.0442 | 0.0179 | 0.0025 | 0.0118 | 0.0019 | 0.0025 | 1.0144 |
| 430 | 93 | 9ix93 | 803DJB93-1-5 | 0.0000 | 0.7283 | 0.0015 | 0.2017 | 0.0475 | 0.0173 | 0.0020 | 0.0084 | 0.0026 | 0.0030 | 1.0123 |
| 431 | 95 | 9ix93 | 803DJB93-2-1 | 0.0000 | 0.6836 | 0.0000 | 0.2077 | 0.0821 | 0.0071 | 0.0044 | 0.0050 | 0.0019 | 0.0165 | 1.0081 |
| 432 | 97 | 9ix93 | 803DJB93-2-2 | 0.0057 | 0.6230 | 0.0003 | 0.1629 | 0.1315 | 0.0363 | 0.0072 | 0.0027 | 0.0029 | 0.0122 | 0.9847 |
| 433 | 98 | 9ix93 | 803DJB93-2-2 | 0.0052 | 0.6147 | 0.0003 | 0.1654 | 0.1400 | 0.0364 | 0.0068 | 0.0011 | 0.0026 | 0.0115 | 0.9840 |
| 434 | 99 | 9ix93 | 803DJB93-2-4 | 0.0078 | 0.6180 | 0.0007 | 0.1557 | 0.1351 | 0.0399 | 0.0064 | 0.0024 | 0.0029 | 0.0118 | 0.9807 |
| 435 | 100 | 9ix93 | 803DJB93-2-4 | 0.0031 | 0.6080 | 0.0004 | 0.1757 | 0.1555 | 0.0392 | 0.0056 | 0.0018 | 0.0027 | 0.0115 | 1.0033 |
| 436 | 101 | 9ix93 | 803DJB93-2-4 | 0.0000 | 0.6175 | 0.0006 | 0.1834 | 0.1378 | 0.0402 | 0.0061 | 0.0018 | 0.0027 | 0.0108 | 1.0008 |
| 437 | 102 | 9ix93 | 803DJB93-2-10 | 0.0073 | 0.6330 | 0.0002 | 0.1704 | 0.1307 | 0.0247 | 0.0072 | 0.0044 | 0.0013 | 0.0053 | 0.9845 |
| 438 | **INDIAN** | | | | | | | | | | | | | |
| 439 | DSDP 761C | | | | | | | | | | | | | |
| 440 | 1 | 9ix93 | 761CJB93-1-1 | 0.0152 | 0.7384 | 0.0006 | 0.1570 | 0.0516 | 0.0269 | 0.0033 | 0.0098 | 0.0021 | 0.0026 | 1.0075 |
| 441 | 2 | 9ix93 | 761CJB93-1-1 | 0.0160 | 0.7312 | 0.0007 | 0.1556 | 0.0555 | 0.0260 | 0.0034 | 0.0087 | 0.0028 | 0.0026 | 1.0024 |
| 442 | 6 | 9ix93 | 761CJB93-1-2 | 0.0165 | 0.7240 | 0.0018 | 0.1480 | 0.0547 | 0.0426 | 0.0022 | 0.0251 | 0.0022 | 0.0025 | 1.0197 |
| 443 | 7 | 9ix93 | 761CJB93-1-2 | 0.0215 | 0.7220 | 0.0011 | 0.1356 | 0.0506 | 0.0407 | 0.0027 | 0.0262 | 0.0024 | 0.0025 | 1.0053 |
| 444 | 8 | 9ix93 | 761CJB93-1-3 | 0.0078 | 0.7242 | 0.0016 | 0.1585 | 0.0400 | 0.0473 | 0.0035 | 0.0124 | 0.0026 | 0.0017 | 0.9997 |
| 445 | 9 | 9ix93 | 761CJB93-1-3 | 0.0081 | 0.7223 | 0.0015 | 0.1594 | 0.0450 | 0.0507 | 0.0032 | 0.0197 | 0.0031 | 0.0016 | 1.0146 |
| 446 | 10 | 9ix93 | 761CJB93-1-3 | 0.0077 | 0.7206 | 0.0016 | 0.1615 | 0.0452 | 0.0459 | 0.0034 | 0.0146 | 0.0033 | 0.0017 | 1.0055 |
| 447 | 11 | 9ix93 | 761CJB93-2-1 | 0.0178 | 0.7381 | 0.0011 | 0.1513 | 0.0448 | 0.0209 | 0.0035 | 0.0077 | 0.0019 | 0.0038 | 0.9909 |
| 448 | 13 | 9ix93 | 761CJB93-2-1 | 0.0152 | 0.7405 | 0.0006 | 0.1585 | 0.0516 | 0.0202 | 0.0048 | 0.0067 | 0.0017 | 0.0035 | 1.0033 |
| 449 | 14 | 9ix93 | 761CJB93-2-2 | 0.0144 | 0.7319 | 0.0013 | 0.1447 | 0.0388 | 0.0461 | 0.0039 | 0.0070 | 0.0039 | 0.0019 | 0.9940 |
| 450 | 15 | 9ix93 | 761CJB93-2-2 | 0.0045 | 0.7311 | 0.0013 | 0.1688 | 0.0440 | 0.0473 | 0.0036 | 0.0094 | 0.0030 | 0.0020 | 1.0151 |
| 451 | 16 | 9ix93 | 761CJB93-2-3 | 0.0227 | 0.7282 | 0.0007 | 0.1327 | 0.0388 | 0.0341 | 0.0042 | 0.0209 | 0.0022 | 0.0020 | 0.9864 |
| 452 | 18 | 9ix93 | 761CJB93-2-3 | 0.0120 | 0.7280 | 0.0012 | 0.1598 | 0.0478 | 0.0347 | 0.0037 | 0.0218 | 0.0019 | 0.0019 | 1.0127 |
| 453 | **ATLANTIC & EUROPE** | | | | | | | | | | | | | |
| 454 | Furlo | | | | | | | | | | | | | |
| 455 | 8 | 24vii84 | D1 | 0.1088 | 0.7104 | 0.0004 | 0.0925 | 0.0118 | 0.0456 | 0.0194 | 0.0034 | 0.0047 | 0.0000 | 0.9970 |
| 456 | 9 | 24vii84 | D1 | 0.0436 | 0.6855 | 0.0000 | 0.1367 | 0.0607 | 0.0773 | 0.0024 | 0.0098 | 0.0050 | 0.0000 | 1.0209 |
| 457 | 18 | 24vii84 | D1 | 0.0914 | 0.7210 | 0.0000 | 0.1112 | 0.0232 | 0.0486 | 0.0140 | 0.0035 | 0.0050 | 0.0000 | 1.0179 |
| 458 | 19 | 24vii84 | D1 | 0.0971 | 0.7194 | 0.0000 | 0.1076 | 0.0198 | 0.0451 | 0.0141 | 0.0029 | 0.0047 | 0.0000 | 1.0106 |
| 459 | 20 | 24vii84 | D1 | 0.0528 | 0.6779 | 0.0004 | 0.1288 | 0.0565 | 0.0666 | 0.0098 | 0.0085 | 0.0046 | 0.0000 | 1.0058 |
| 460 | 21 | 24vii84 | R1 | 0.1387 | 0.6907 | 0.0000 | 0.0832 | 0.0281 | 0.0467 | 0.0139 | 0.0069 | 0.0072 | 0.0000 | 1.0154 |
| 461 | 23 | 24vii84 | R1 | 0.1303 | 0.6742 | 0.0003 | 0.0864 | 0.0292 | 0.0519 | 0.0115 | 0.0100 | 0.0101 | 0.0000 | 1.0039 |
| 462 | 25 | 24vii84 | P1 | 0.1119 | 0.7118 | 0.0000 | 0.0954 | 0.0249 | 0.0545 | 0.0115 | 0.0026 | 0.0043 | 0.0000 | 1.0169 |
| 463 | 27 | 24vii84 | P1 | 0.1235 | 0.6842 | 0.0000 | 0.0835 | 0.0322 | 0.0566 | 0.0149 | 0.0068 | 0.0049 | 0.0000 | 1.0066 |
| 464 | 28 | 24vii84 | P1 | 0.1030 | 0.6924 | 0.0004 | 0.1045 | 0.0414 | 0.0553 | 0.0094 | 0.0055 | 0.0053 | 0.0000 | 1.0171 |
| 465 | 32 | 24vii84 | P1 | 0.0772 | 0.6827 | 0.0001 | 0.1250 | 0.0478 | 0.0483 | 0.0074 | 0.0074 | 0.0058 | 0.0000 | 1.0016 |
| 466 | 33 | 24vii84 | P1 | 0.0672 | 0.6995 | 0.0000 | 0.1270 | 0.0466 | 0.0602 | 0.0068 | 0.0072 | 0.0047 | 0.0000 | 1.0192 |
| 467 | 34 | 24vii84 | P1 | 0.0797 | 0.6957 | 0.0001 | 0.1243 | 0.0495 | 0.0508 | 0.0076 | 0.0045 | 0.0048 | 0.0000 | 1.0168 |
| 468 | 35 | 24vii84 | P1 | 0.1057 | 0.6565 | 0.0000 | 0.0991 | 0.0557 | 0.0609 | 0.0076 | 0.0160 | 0.0040 | 0.0000 | 1.0055 |
| 469 | 36 | 24vii84 | G2 | 0.0575 | 0.7129 | 0.0004 | 0.1521 | 0.0497 | 0.0298 | 0.0099 | 0.0009 | 0.0067 | 0.0000 | 1.0199 |
| 470 | 37 | 24vii84 | G2 | 0.0521 | 0.6848 | 0.0004 | 0.1502 | 0.0593 | 0.0416 | 0.0071 | 0.0175 | 0.0055 | 0.0000 | 1.0186 |
| 471 | 38 | 24vii84 | G2 | 0.0502 | 0.7025 | 0.0007 | 0.1527 | 0.0508 | 0.0373 | 0.0091 | 0.0082 | 0.0070 | 0.0000 | 1.0185 |
| 472 | 39 | 24vii84 | G2 | 0.0659 | 0.7053 | 0.0000 | 0.1420 | 0.0415 | 0.0283 | 0.0107 | 0.0023 | 0.0057 | 0.0000 | 1.0018 |
| 473 | 40 | 24vii84 | G2 | 0.0797 | 0.7144 | 0.0000 | 0.1297 | 0.0260 | 0.0206 | 0.0177 | 0.0007 | 0.0052 | 0.0000 | 0.9940 |
| 474 | 41 | 24vii84 | G2 | 0.0784 | 0.7203 | 0.0000 | 0.1320 | 0.0282 | 0.0220 | 0.0170 | 0.0010 | 0.0047 | 0.0000 | 1.0035 |
| 475 | 42 | 24vii84 | G2 | 0.0692 | 0.7156 | 0.0000 | 0.1385 | 0.0339 | 0.0252 | 0.0159 | 0.0007 | 0.0060 | 0.0000 | 1.0049 |
| 476 | 43 | 24vii84 | G2 | 0.0552 | 0.6913 | 0.0000 | 0.1490 | 0.0498 | 0.0288 | 0.0082 | 0.0024 | 0.0058 | 0.0000 | 0.9905 |
| 477 | 3 | 9viii84 | P2 | 0.1194 | 0.6883 | 0.0000 | 0.0750 | 0.0158 | 0.0460 | 0.0255 | 0.0022 | 0.0050 | 0.0053 | 0.9825 |
| 478 | 4 | 9viii84 | P2 | 0.0707 | 0.6915 | 0.0002 | 0.1171 | 0.0432 | 0.0571 | 0.0137 | 0.0041 | 0.0060 | 0.0040 | 1.0074 |
| 479 | 5 | 9viii84 | P2 | 0.0662 | 0.6881 | 0.0000 | 0.1241 | 0.0467 | 0.0571 | 0.0084 | 0.0026 | 0.0076 | 0.0045 | 1.0054 |
| 480 | 6 | 9viii84 | P2 | 0.0241 | 0.6717 | 0.0004 | 0.1792 | 0.0649 | 0.0221 | 0.0024 | 0.0357 | 0.0040 | 0.0033 | 1.0077 |
| 481 | 7 | 9viii84 | P2 | 0.1133 | 0.6888 | 0.0000 | 0.0915 | 0.0263 | 0.0392 | 0.0173 | 0.0028 | 0.0052 | 0.0052 | 0.9895 |
| 482 | 8 | 9viii84 | P2 | 0.0642 | 0.6990 | 0.0000 | 0.1216 | 0.0348 | 0.0553 | 0.0109 | 0.0048 | 0.0060 | 0.0047 | 1.0013 |
| 483 | 10 | 9viii84 | P2 | 0.0548 | 0.6956 | 0.0000 | 0.1325 | 0.0500 | 0.0573 | 0.0080 | 0.0019 | 0.0059 | 0.0036 | 1.0098 |
| 484 | 11 | 9viii84 | P2 | 0.0899 | 0.6772 | 0.0002 | 0.1055 | 0.0432 | 0.0498 | 0.0127 | 0.0034 | 0.0047 | 0.0043 | 0.9909 |
| 485 | 12 | 9viii84 | P2 | 0.0420 | 0.6586 | 0.0003 | 0.1465 | 0.0441 | 0.0486 | 0.0036 | 0.0550 | 0.0023 | 0.0025 | 1.0034 |
| 486 | 14 | 9viii84 | F1 | 0.1352 | 0.6778 | 0.0000 | 0.0487 | 0.0124 | 0.0794 | 0.0170 | 0.0044 | 0.0042 | 0.0053 | 0.9844 |
| 487 | 15 | 9viii84 | F1 | 0.1035 | 0.6761 | 0.0000 | 0.0654 | 0.0165 | 0.0951 | 0.0124 | 0.0119 | 0.0050 | 0.0043 | 0.9902 |
| 488 | 17 | 9viii84 | P3 | 0.0424 | 0.6938 | 0.0000 | 0.1284 | 0.0475 | 0.0735 | 0.0083 | 0.0036 | 0.0045 | 0.0034 | 1.0054 |
| 489 | 18 | 9viii84 | P3 | 0.0739 | 0.7158 | 0.0000 | 0.1174 | 0.0166 | 0.0248 | 0.0241 | 0.0027 | 0.0035 | 0.0072 | 0.9859 |
| 490 | 19 | 9viii84 | P3 | 0.0828 | 0.6839 | 0.0003 | 0.1032 | 0.0324 | 0.0532 | 0.0149 | 0.0026 | 0.0036 | 0.0034 | 0.9805 |
| 491 | 20 | 9viii84 | P3 | 0.0527 | 0.6829 | 0.0001 | 0.1234 | 0.0474 | 0.0710 | 0.0068 | 0.0083 | 0.0043 | 0.0031 | 0.9998 |
| 492 | 21 | 9viii84 | P3 | 0.1350 | 0.6818 | 0.0003 | 0.0706 | 0.0216 | 0.0466 | 0.0183 | 0.0025 | 0.0043 | 0.0046 | 0.9855 |
| 493 | 25 | 9viii84 | F2 | 0.1370 | 0.6345 | 0.0002 | 0.0611 | 0.0260 | 0.0748 | 0.0108 | 0.0378 | 0.0060 | 0.0023 | 0.9905 |
| 494 | 26 | 9viii84 | G3 | 0.1075 | 0.6994 | 0.0000 | 0.0933 | 0.0194 | 0.0378 | 0.0171 | 0.0017 | 0.0045 | 0.0071 | 0.9877 |
| 495 | 27 | 9viii84 | G3 | 0.1010 | 0.7121 | 0.0000 | 0.1068 | 0.0189 | 0.0295 | 0.0152 | 0.0011 | 0.0054 | 0.0061 | 0.9963 |
| 496 | 29 | 9viii84 | G3 | 0.0797 | 0.7018 | 0.0000 | 0.1184 | 0.0374 | 0.0408 | 0.0134 | 0.0008 | 0.0053 | 0.0060 | 1.0035 |
| 497 | 30 | 9viii84 | G3 | 0.0503 | 0.7164 | 0.0003 | 0.1464 | 0.0448 | 0.0359 | 0.0100 | 0.0017 | 0.0053 | 0.0057 | 1.0169 |
| 498 | 31 | 9viii84 | G3 | 0.1020 | 0.6976 | 0.0002 | 0.0913 | 0.0247 | 0.0485 | 0.0186 | 0.0011 | 0.0047 | 0.0071 | 0.9958 |
| 499 | 13 | 23viii84 | P4 | 0.1738 | 0.6655 | 0.0000 | 0.0372 | 0.0159 | 0.0671 | 0.0145 | 0.0138 | 0.0048 | 0.0045 | 0.9970 |
| 500 | 14 | 23viii84 | P4 | 0.1709 | 0.6776 | 0.0000 | 0.0440 | 0.0139 | 0.0619 | 0.0123 | 0.0084 | 0.0054 | 0.0047 | 0.9990 |
| 501 | 15 | 23viii84 | P4 | 0.1604 | 0.6761 | 0.0000 | 0.0514 | 0.0172 | 0.0642 | 0.0108 | 0.0091 | 0.0054 | 0.0039 | 0.9984 |
| 502 | 16 | 23viii84 | P4 | 0.1390 | 0.6724 | 0.0000 | 0.0617 | 0.0233 | 0.0740 | 0.0096 | 0.0146 | 0.0051 | 0.0034 | 1.0030 |
| 503 | 17 | 23viii84 | P4 | 0.1763 | 0.6652 | 0.0000 | 0.0428 | 0.0193 | 0.0580 | 0.0123 | 0.0097 | 0.0050 | 0.0039 | 0.9924 |
| 504 | 18 | 23viii84 | P4 | 0.1469 | 0.6842 | 0.0000 | 0.0665 | 0.0201 | 0.0530 | 0.0115 | 0.0059 | 0.0047 | 0.0035 | 0.9962 |
| 505 | 19 | 23viii84 | P4 | 0.1865 | 0.6751 | 0.0000 | 0.0443 | 0.0176 | 0.0496 | 0.0117 | 0.0060 | 0.0063 | 0.0035 | 1.0006 |
| 506 | 20 | 23viii84 | P4 | 0.1755 | 0.6601 | 0.0000 | 0.0175 | 0.0129 | 0.0923 | 0.0156 | 0.0103 | 0.0037 | 0.0043 | 0.9922 |
| 507 | 24 | 23viii84 | P5 | 0.0830 | 0.7185 | 0.0000 | 0.1280 | 0.0374 | 0.0306 | 0.0100 | 0.0025 | 0.0043 | 0.0055 | 1.0198 |
| 508 | 25 | 23viii84 | P5 | 0.0911 | 0.7151 | 0.0000 | 0.1224 | 0.0378 | 0.0300 | 0.0111 | 0.0019 | 0.0043 | 0.0052 | 1.0189 |
| 509 | 26 | 23viii84 | P5 | 0.0623 | 0.7025 | 0.0000 | 0.1368 | 0.0487 | 0.0427 | 0.0073 | 0.0028 | 0.0049 | 0.0038 | 1.0118 |
| 510 | 27 | 23viii84 | P5 | 0.0271 | 0.6544 | 0.0004 | 0.1676 | 0.0888 | 0.0436 | 0.0029 | 0.0206 | 0.0043 | 0.0024 | 1.0120 |
| 511 | 28 | 23viii84 | P5 | 0.0537 | 0.6549 | 0.0002 | 0.1430 | 0.0892 | 0.0525 | 0.0046 | 0.0084 | 0.0039 | 0.0029 | 1.0135 |
| 512 | 29 | 23viii84 | P5 | 0.0837 | 0.6925 | 0.0000 | 0.1220 | 0.0464 | 0.0377 | 0.0105 | 0.0047 | 0.0047 | 0.0046 | 1.0068 |
| 513 | 30 | 23viii84 | P5 | 0.0730 | 0.7007 | 0.0000 | 0.1285 | 0.0435 | 0.0374 | 0.0114 | 0.0071 | 0.0041 | 0.0049 | 1.0106 |
| 514 | 31 | 23viii84 | P5 | 0.0582 | 0.6853 | 0.0000 | 0.1380 | 0.0599 | 0.0450 | 0.0075 | 0.0032 | 0.0049 | 0.0039 | 1.0058 |
| 515 | 32 | 23viii84 | P5 | 0.0424 | 0.6572 | 0.0001 | 0.1472 | 0.0801 | 0.0516 | 0.0036 | 0.0075 | 0.0037 | 0.0028 | 0.9962 |
| 516 | 33 | 23viii84 | P6 | 0.1952 | 0.6867 | 0.0001 | 0.0378 | 0.0043 | 0.0434 | 0.0155 | 0.0023 | 0.0061 | 0.0024 | 0.9939 |
| 517 | 34 | 23viii84 | P6 | 0.2878 | 0.6635 | 0.0002 | 0.0073 | 0.0079 | 0.0082 | 0.0060 | 0.0016 | 0.0066 | 0.0010 | 0.9901 |
| 518 | 35 | 23viii84 | P6 | 0.1563 | 0.6851 | 0.0000 | 0.0502 | 0.0119 | 0.0571 | 0.0223 | 0.0013 | 0.0061 | 0.0033 | 0.9936 |
| 519 | 36 | 23viii84 | P6 | 0.1645 | 0.6896 | 0.0000 | 0.0522 | 0.0176 | 0.0554 | 0.0160 | 0.0033 | 0.0044 | 0.0034 | 1.0063 |
| 520 | 38 | 23viii84 | F3 | 0.2454 | 0.6592 | 0.0000 | 0.0017 | 0.0052 | 0.0431 | 0.0183 | 0.0046 | 0.0056 | 0.0024 | 0.9855 |
| 521 | 39 | 23viii84 | F3 | 0.3198 | 0.5872 | 0.0000 | 0.0001 | 0.0050 | 0.0085 | 0.0040 | 0.0035 | 0.0356 | 0.0005 | 0.9641 |
| 522 | 40 | 23viii84 | F3 | 0.2453 | 0.6740 | 0.0000 | 0.0119 | 0.0049 | 0.0268 | 0.0199 | 0.0043 | 0.0051 | 0.0037 | 0.9958 |
| 523 | 41 | 23viii84 | F4 | 0.0512 | 0.7088 | 0.0000 | 0.1129 | 0.0294 | 0.0793 | 0.0160 | 0.0054 | 0.0050 | 0.0040 | 1.0121 |
| 524 | 42 | 23viii84 | F4 | 0.0820 | 0.7141 | 0.0000 | 0.1049 | 0.0246 | 0.0562 | 0.0162 | 0.0038 | 0.0037 | 0.0054 | 1.0108 |
| 525 | 43 | 23viii84 | F4 | 0.0471 | 0.6807 | 0.0002 | 0.1009 | 0.0332 | 0.1017 | 0.0111 | 0.0042 | 0.0048 | 0.0026 | 0.9867 |
| 526 | 44 | 23viii84 | F5 | 0.0689 | 0.7009 | 0.0000 | 0.1171 | 0.0360 | 0.0611 | 0.0081 | 0.0042 | 0.0053 | 0.0053 | 1.0069 |
| 527 | 45 | 23viii84 | F5 | 0.1834 | 0.6706 | 0.0001 | 0.0178 | 0.0045 | 0.0755 | 0.0222 | 0.0044 | 0.0052 | 0.0041 | 0.9878 |
| 528 | 46 | 23viii84 | F5 | 0.1234 | 0.6924 | 0.0000 | 0.0634 | 0.0174 | 0.0749 | 0.0168 | 0.0064 | 0.0046 | 0.0067 | 1.0061 |
| 529 | 10 | 30viii84 | D2 | 0.0560 | 0.7113 | 0.0006 | 0.1338 | 0.0270 | 0.0505 | 0.0090 | 0.0057 | 0.0067 | 0.0006 | 1.0011 |
| 530 | 11 | 30viii84 | D2 | 0.0537 | 0.7106 | 0.0004 | 0.1321 | 0.0281 | 0.0545 | 0.0083 | 0.0042 | 0.0061 | 0.0012 | 0.9993 |
| 531 | 13 | 30viii84 | D2 | 0.0556 | 0.7110 | 0.0002 | 0.1326 | 0.0251 | 0.0496 | 0.0097 | 0.0054 | 0.0063 | 0.0009 | 0.9965 |
| 532 | 17 | 30viii84 | D2 | 0.0478 | 0.7111 | 0.0001 | 0.1402 | 0.0271 | 0.0454 | 0.0102 | 0.0069 | 0.0063 | 0.0008 | 0.9958 |
| 533 | 19 | 30viii84 | P7 | 0.1164 | 0.6578 | 0.0000 | 0.0516 | 0.0181 | 0.0884 | 0.0143 | 0.0067 | 0.0018 | 0.0049 | 0.9600 |
| 534 | 25 | 30viii84 | P7 | 0.0958 | 0.6744 | 0.0000 | 0.0717 | 0.0267 | 0.0924 | 0.0097 | 0.0068 | 0.0022 | 0.0047 | 0.9845 |
| 535 | 26 | 30viii84 | P7 | 0.0904 | 0.6671 | 0.0000 | 0.0933 | 0.0511 | 0.0794 | 0.0079 | 0.0061 | 0.0035 | 0.0027 | 1.0016 |
| 536 | 30 | 30viii84 | P8 | 0.0415 | 0.6700 | 0.0000 | 0.1462 | 0.0769 | 0.0579 | 0.0053 | 0.0055 | 0.0053 | 0.0030 | 1.0116 |
| 537 | 31 | 30viii84 | P8 | 0.0714 | 0.6915 | 0.0000 | 0.1145 | 0.0349 | 0.0514 | 0.0118 | 0.0014 | 0.0045 | 0.0053 | 0.9868 |
| 538 | 32 | 30viii84 | P8 | 0.0757 | 0.6915 | 0.0000 | 0.1161 | 0.0442 | 0.0512 | 0.0117 | 0.0016 | 0.0040 | 0.0045 | 1.0005 |
| 539 | 33 | 30viii84 | P8 | 0.0672 | 0.6777 | 0.0000 | 0.1145 | 0.0495 | 0.0680 | 0.0100 | 0.0075 | 0.0052 | 0.0039 | 1.0035 |
| 540 | 34 | 30viii84 | P8 | 0.0304 | 0.6769 | 0.0007 | 0.1712 | 0.0749 | 0.0299 | 0.0036 | 0.0168 | 0.0037 | 0.0024 | 1.0106 |
| 541 | 36 | 30viii84 | P8 | 0.0859 | 0.6875 | 0.0000 | 0.1041 | 0.0388 | 0.0559 | 0.0137 | 0.0022 | 0.0047 | 0.0049 | 0.9976 |
| 542 | 37 | 30viii84 | P8 | 0.0429 | 0.6747 | 0.0004 | 0.1541 | 0.0698 | 0.0374 | 0.0073 | 0.0040 | 0.0058 | 0.0026 | 0.9990 |
| 543 | 39 | 30viii84 | P9 | 0.0699 | 0.6959 | 0.0000 | 0.1064 | 0.0223 | 0.0536 | 0.0175 | 0.0033 | 0.0042 | 0.0063 | 0.9794 |
| 544 | 42 | 30viii84 | P9 | 0.0205 | 0.6303 | 0.0005 | 0.1686 | 0.1021 | 0.0402 | 0.0038 | 0.0123 | 0.0035 | 0.0024 | 0.9842 |
| 545 | 6 | 13ix84 | P10 | 0.1703 | 0.6719 | 0.0000 | 0.0469 | 0.0232 | 0.0668 | 0.0100 | 0.0070 | 0.0054 | 0.0021 | 1.0037 |
| 546 | 7 | 13ix84 | P10 | 0.1698 | 0.6707 | 0.0000 | 0.0473 | 0.0223 | 0.0655 | 0.0103 | 0.0089 | 0.0052 | 0.0020 | 1.0021 |
| 547 | 8 | 13ix84 | P10 | 0.1698 | 0.6690 | 0.0000 | 0.0492 | 0.0233 | 0.0611 | 0.0107 | 0.0083 | 0.0053 | 0.0022 | 0.9989 |
| 548 | 9 | 13ix84 | P10 | 0.1008 | 0.6704 | 0.0000 | 0.0933 | 0.0382 | 0.0716 | 0.0056 | 0.0111 | 0.0060 | 0.0016 | 0.9985 |
| 549 | 11 | 13ix84 | P10 | 0.1490 | 0.6851 | 0.0000 | 0.0703 | 0.0223 | 0.0400 | 0.0175 | 0.0042 | 0.0049 | 0.0026 | 0.9958 |
| 550 | 12 | 13ix84 | P10 | 0.0914 | 0.6715 | 0.0002 | 0.1006 | 0.0365 | 0.0638 | 0.0080 | 0.0138 | 0.0050 | 0.0020 | 0.9927 |
| 551 | 23 | 13ix84 | P11 | 0.0698 | 0.6995 | 0.0000 | 0.1291 | 0.0350 | 0.0327 | 0.0114 | 0.0075 | 0.0044 | 0.0056 | 0.9949 |
| 552 | 24 | 13ix84 | P11 | 0.0712 | 0.7085 | 0.0000 | 0.1285 | 0.0349 | 0.0344 | 0.0126 | 0.0067 | 0.0039 | 0.0052 | 1.0059 |
| 553 | 26 | 13ix84 | P11 | 0.0600 | 0.7089 | 0.0000 | 0.1271 | 0.0301 | 0.0456 | 0.0123 | 0.0064 | 0.0038 | 0.0051 | 0.9992 |
| 554 | 27 | 13ix84 | P11 | 0.0623 | 0.7016 | 0.0002 | 0.1139 | 0.0244 | 0.0617 | 0.0135 | 0.0053 | 0.0053 | 0.0045 | 0.9927 |
| 555 | 28 | 13ix84 | P11 | 0.0681 | 0.7009 | 0.0000 | 0.1189 | 0.0300 | 0.0492 | 0.0125 | 0.0034 | 0.0051 | 0.0051 | 0.9931 |
| 556 | 29 | 13ix84 | P12 | 0.0485 | 0.6894 | 0.0000 | 0.1338 | 0.0486 | 0.0572 | 0.0074 | 0.0047 | 0.0050 | 0.0042 | 0.9989 |
| 557 | 30 | 13ix84 | P12 | 0.0638 | 0.7003 | 0.0001 | 0.1237 | 0.0373 | 0.0562 | 0.0070 | 0.0029 | 0.0051 | 0.0043 | 1.0006 |
| 558 | 31 | 13ix84 | P12 | 0.1222 | 0.6728 | 0.0000 | 0.0862 | 0.0389 | 0.0526 | 0.0097 | 0.0087 | 0.0044 | 0.0039 | 0.9995 |
| 559 | 32 | 13ix84 | P12 | 0.0621 | 0.6891 | 0.0000 | 0.1251 | 0.0446 | 0.0580 | 0.0066 | 0.0058 | 0.0051 | 0.0039 | 1.0003 |
| 560 | 35 | 13ix84 | P13 | 0.1684 | 0.6713 | 0.0000 | 0.0402 | 0.0123 | 0.0598 | 0.0163 | 0.0056 | 0.0045 | 0.0051 | 0.9836 |
| 561 | 40 | 13ix84 | O2 | 0.0812 | 0.6911 | 0.0000 | 0.1110 | 0.0324 | 0.0543 | 0.0089 | 0.0023 | 0.0054 | 0.0028 | 0.9893 |
| 562 | 41 | 13ix84 | P14 | 0.1390 | 0.6855 | 0.0000 | 0.0561 | 0.0128 | 0.0667 | 0.0127 | 0.0021 | 0.0033 | 0.0059 | 0.9841 |
| 563 | 42 | 13ix84 | P14 | 0.0954 | 0.6719 | 0.0000 | 0.0869 | 0.0372 | 0.0745 | 0.0080 | 0.0050 | 0.0030 | 0.0043 | 0.9861 |
| 564 | 43 | 13ix84 | P14 | 0.1173 | 0.6693 | 0.0000 | 0.0656 | 0.0265 | 0.0810 | 0.0110 | 0.0112 | 0.0036 | 0.0057 | 0.9912 |
| 565 | 43 | 13ix84 | O3 | 0.1896 | 0.6701 | 0.0002 | 0.0464 | 0.0135 | 0.0404 | 0.0105 | 0.0021 | 0.0073 | 0.0018 | 0.9820 |
| 566 | 45 | 13ix84 | O3 | 0.1047 | 0.6679 | 0.0000 | 0.0908 | 0.0380 | 0.0593 | 0.0100 | 0.0051 | 0.0049 | 0.0030 | 0.9837 |
| 567 | 46 | 13ix84 | O3 | 0.1656 | 0.6769 | 0.0000 | 0.0548 | 0.0088 | 0.0376 | 0.0151 | 0.0024 | 0.0042 | 0.0030 | 0.9683 |
| 568 | 51 | 13ix84 | P15 | 0.1307 | 0.6635 | 0.0000 | 0.0657 | 0.0235 | 0.0656 | 0.0157 | 0.0105 | 0.0050 | 0.0010 | 0.9811 |

| | | | | | | | | | | | | | |
|---|---|---|---|---|---|---|---|---|---|---|---|---|---|
| 569 | 52 | 13ix84 | P15 | 0.1189 | 0.6691 | 0.0000 | 0.0750 | 0.0257 | 0.0653 | 0.0138 | 0.0065 | 0.0052 | 0.0011 | 0.9806 |
| 570 | 53 | 13ix84 | P15 | 0.1335 | 0.6593 | 0.0002 | 0.0666 | 0.0262 | 0.0655 | 0.0119 | 0.0115 | 0.0050 | 0.0012 | 0.9808 |
| 571 | 57 | 13ix84 | P16 | 0.0715 | 0.6890 | 0.0000 | 0.0799 | 0.0175 | 0.1000 | 0.0138 | 0.0046 | 0.0033 | 0.0043 | 0.9840 |
| 572 | 58 | 13ix84 | P16 | 0.0692 | 0.6876 | 0.0000 | 0.0822 | 0.0172 | 0.0977 | 0.0144 | 0.0044 | 0.0038 | 0.0040 | 0.9805 |
| 573 | 60 | 13ix84 | P16 | 0.0509 | 0.6677 | 0.0000 | 0.1162 | 0.0497 | 0.0799 | 0.0069 | 0.0079 | 0.0044 | 0.0035 | 0.9870 |
| 574 | 61 | 13ix84 | P16 | 0.0630 | 0.6874 | 0.0001 | 0.0921 | 0.0255 | 0.0898 | 0.0129 | 0.0038 | 0.0036 | 0.0044 | 0.9826 |
| 575 | 62 | 13ix84 | P16 | 0.0489 | 0.6787 | 0.0000 | 0.1210 | 0.0479 | 0.0764 | 0.0058 | 0.0052 | 0.0042 | 0.0034 | 0.9915 |
| 576 | 63 | 13ix84 | P16 | 0.0809 | 0.6851 | 0.0000 | 0.0791 | 0.0231 | 0.0952 | 0.0127 | 0.0028 | 0.0039 | 0.0044 | 0.9872 |
| 577 | 64 | 13ix84 | P16 | 0.0487 | 0.6765 | 0.0000 | 0.1244 | 0.0542 | 0.0700 | 0.0073 | 0.0055 | 0.0036 | 0.0043 | 0.9946 |
| 578 | 65 | 13ix84 | P16 | 0.0733 | 0.6865 | 0.0006 | 0.0984 | 0.0240 | 0.0624 | 0.0154 | 0.0017 | 0.0032 | 0.0057 | 0.9712 |
| 579 | 66 | 13ix84 | P16 | 0.0631 | 0.6804 | 0.0000 | 0.1051 | 0.0362 | 0.0702 | 0.0130 | 0.0026 | 0.0037 | 0.0059 | 0.9802 |
| 580 | 67 | 13ix84 | P16 | 0.0451 | 0.6796 | 0.0001 | 0.1254 | 0.0529 | 0.0726 | 0.0068 | 0.0025 | 0.0038 | 0.0038 | 0.9926 |
| 581 | DSDP 524 | | | | | | | | | | | | | |
| 582 | 20 | old | DSDP524 | 0.1421 | 0.6786 | 0.0003 | 0.0666 | 0.0158 | 0.0342 | 0.0216 | 0.0069 | 0.0027 | 0.0064 | 0.9752 |
| 583 | 21 | old | DSDP524 | 0.0715 | 0.6778 | 0.0003 | 0.1113 | 0.0408 | 0.0611 | 0.0101 | 0.0062 | 0.0047 | 0.0041 | 0.9879 |
| 584 | 22 | old | DSDP524 | 0.1386 | 0.6691 | 0.0003 | 0.0747 | 0.0252 | 0.0423 | 0.0140 | 0.0055 | 0.0058 | 0.0035 | 0.9790 |
| 585 | 23 | old | DSDP524 | 0.1666 | 0.6707 | 0.0001 | 0.0435 | 0.0082 | 0.0449 | 0.0198 | 0.0055 | 0.0026 | 0.0054 | 0.9673 |
| 586 | 7 | old | DSDP524 | 0.1138 | 0.6071 | 0.0138 | 0.0782 | 0.0238 | 0.0430 | 0.0136 | 0.0179 | 0.0075 | 0.0021 | 0.9208 |
| 587 | 9 | old | DSDP524 | 0.1843 | 0.5872 | 0.0226 | 0.0478 | 0.0179 | 0.0119 | 0.0105 | 0.0043 | 0.0095 | 0.0029 | 0.8988 |
| 588 | 10 | old | DSDP524 | 0.1686 | 0.5809 | 0.0255 | 0.0494 | 0.0165 | 0.0184 | 0.0087 | 0.0040 | 0.0077 | 0.0037 | 0.8834 |
| 589 | Caravaca | | | | | | | | | | | | | |
| 590 | 23 | 30ix93 | CaraJB93-2-1 | 0.0788 | 0.5760 | 0.0012 | 0.1364 | 0.0683 | 0.0373 | 0.0019 | 0.0991 | 0.0029 | 0.0018 | 1.0038 |
| 591 | 24 | 30ix93 | CaraJB93-2-1 | 0.0794 | 0.5910 | 0.0011 | 0.1368 | 0.0718 | 0.0411 | 0.0028 | 0.0916 | 0.0026 | 0.0013 | 1.0196 |
| 592 | 25 | 30ix93 | CaraJB93-2-2 | 0.0687 | 0.5630 | 0.0007 | 0.1485 | 0.0695 | 0.0302 | 0.0020 | 0.1215 | 0.0025 | 0.0012 | 1.0078 |
| 593 | 26 | 30ix93 | CaraJB93-2-2 | 0.0783 | 0.5602 | 0.0009 | 0.1421 | 0.0697 | 0.0292 | 0.0021 | 0.1162 | 0.0029 | 0.0012 | 1.0028 |
| 594 | 27 | 30ix93 | CaraJB93-2-3 | 0.0563 | 0.5331 | 0.0009 | 0.1524 | 0.0543 | 0.0345 | 0.0025 | 0.1676 | 0.0029 | 0.0009 | 1.0052 |
| 595 | 28 | 30ix93 | CaraJB93-2-3 | 0.0919 | 0.5137 | 0.0006 | 0.1267 | 0.0586 | 0.0316 | 0.0021 | 0.1525 | 0.0031 | 0.0009 | 0.9817 |
| 596 | 29 | 30ix93 | CaraJB93-2-4 | 0.0847 | 0.5101 | 0.0010 | 0.1347 | 0.0827 | 0.0326 | 0.0016 | 0.1337 | 0.0030 | 0.0008 | 0.9851 |
| 597 | 30 | 30ix93 | CaraJB93-2-4 | 0.0238 | 0.5469 | 0.0011 | 0.1767 | 0.0902 | 0.0344 | 0.0020 | 0.1202 | 0.0030 | 0.0005 | 0.9989 |
| 598 | 34 | 30ix93 | CaraJB93-2-5b | 0.0887 | 0.6084 | 0.0009 | 0.1185 | 0.0537 | 0.0472 | 0.0023 | 0.0616 | 0.0034 | 0.0010 | 0.9857 |
| 599 | 35 | 30ix93 | CaraJB93-2-? | 0.0811 | 0.3882 | 0.0008 | 0.1385 | 0.0887 | 0.0319 | 0.0021 | 0.2493 | 0.0028 | 0.0010 | 0.9843 |
| 600 | 36 | 30ix93 | CaraJB93-2-9 | 0.0935 | 0.6127 | 0.0009 | 0.1262 | 0.0613 | 0.0333 | 0.0041 | 0.0586 | 0.0039 | 0.0010 | 0.9956 |
| 601 | 37 | 30ix93 | CaraJB93-2-9 | 0.1048 | 0.6073 | 0.0013 | 0.1182 | 0.0623 | 0.0361 | 0.0032 | 0.0567 | 0.0041 | 0.0008 | 0.9946 |
| 602 | 38 | 30ix93 | CaraJB93-2-10 | 0.2009 | 0.5746 | 0.0011 | 0.0552 | 0.0417 | 0.0341 | 0.0052 | 0.0696 | 0.0082 | 0.0008 | 0.9912 |
| 603 | 39 | 30ix93 | CaraJB93-2-10 | 0.1998 | 0.5806 | 0.0006 | 0.0532 | 0.0426 | 0.0371 | 0.0043 | 0.0645 | 0.0071 | 0.0010 | 0.9908 |
| 604 | 40 | 30ix93 | CaraJB93-2-12 | 0.1006 | 0.4965 | 0.0016 | 0.1244 | 0.0704 | 0.0310 | 0.0024 | 0.1572 | 0.0025 | 0.0010 | 0.9876 |
| 605 | 41 | 30ix93 | CaraJB93-2-12 | 0.1014 | 0.5974 | 0.0006 | 0.1056 | 0.0721 | 0.0584 | 0.0026 | 0.0477 | 0.0034 | 0.0009 | 0.9901 |
| 606 | 42 | 30ix93 | CaraJB93-2-12 | 0.0797 | 0.6307 | 0.0008 | 0.1188 | 0.0591 | 0.0569 | 0.0034 | 0.0387 | 0.0031 | 0.0013 | 0.9925 |
| 607 | 43 | 30ix93 | CaraJB93-2-11 | 0.0853 | 0.4930 | 0.0011 | 0.1358 | 0.0735 | 0.0303 | 0.0028 | 0.1688 | 0.0027 | 0.0013 | 0.9946 |
| 608 | 44 | 30ix93 | CaraJB93-2-11 | 0.0896 | 0.5102 | 0.0005 | 0.1300 | 0.0680 | 0.0302 | 0.0021 | 0.1457 | 0.0033 | 0.0010 | 0.9805 |

Ebel and Grossman (2005) *Geology*
Data Repository Table DR2:

**Table DR2:  Abundances of elements (atoms) in initial vapor plume.**

|    | 45C | 45S | 60C | 60S | 90C | 90S | 60S+15%air |
|----|-----|-----|-----|-----|-----|-----|------------|
| H  | 0.171460559 | 0.148914173 | 0.144995797 | 0.123515533 | 0.17848487 | 0.151966023 | 0.10291941 |
| C  | 0.106987458 | 0.092476603 | 0.097277821 | 0.08250522 | 0.081449725 | 0.068982311 | 0.068772724 |
| N  | 2.72459E-06 | 3.29472E-06 | 7.27428E-06 | 8.61027E-06 | 8.74777E-06 | 1.03482E-05 | 0.131424605 |
| O  | 0.53210378 | 0.545686715 | 0.530127627 | 0.540177045 | 0.512118245 | 0.52277659 | 0.485409736 |
| Na | 0.002522315 | 0.002810047 | 0.003955252 | 0.004484151 | 0.004904222 | 0.005557937 | 0.003736419 |
| Mg | 0.024049378 | 0.022564403 | 0.03071493 | 0.030592722 | 0.030707444 | 0.031483858 | 0.025491388 |
| Al | 0.00572217 | 0.006715923 | 0.009740084 | 0.011341214 | 0.011626332 | 0.013594846 | 0.009450068 |
| Si | 0.028598239 | 0.033144146 | 0.048586933 | 0.056184349 | 0.056852248 | 0.066133471 | 0.046815612 |
| P  | 3.37168E-05 | 4.07722E-05 | 9.00192E-05 | 0.000106552 | 0.000108254 | 0.000128059 | 8.87846E-05 |
| S  | 0.01484766 | 0.02927222 | 0.01391197 | 0.025852921 | 0.012283699 | 0.022300423 | 0.021541948 |
| K  | 0.001267366 | 0.001527488 | 0.002266204 | 0.002678236 | 0.002818319 | 0.003328795 | 0.00223164 |
| Ca | 0.103194282 | 0.105706562 | 0.09419476 | 0.093990169 | 0.079579501 | 0.07936254 | 0.078317314 |
| Ti | 0.00011054 | 0.000133671 | 0.000227377 | 0.000269138 | 0.000279191 | 0.000330271 | 0.000224259 |
| Cr | 0.000122834 | 0.000148538 | 0.000327751 | 0.000387945 | 0.000394157 | 0.000466271 | 0.000323256 |
| Mn | 5.67803E-05 | 6.86618E-05 | 0.000145101 | 0.00017175 | 0.000175045 | 0.00020707 | 0.000143111 |
| Fe | 0.008441467 | 0.010207876 | 0.022153002 | 0.026221614 | 0.026673008 | 0.031552993 | 0.021849161 |
| Co | 2.23184E-05 | 2.69887E-05 | 5.95776E-05 | 7.05196E-05 | 7.16465E-05 | 8.47547E-05 | 5.87605E-05 |
| Ni | 0.000456412 | 0.000551917 | 0.001218518 | 0.001442311 | 0.001465346 | 0.001733439 | 0.001201805 |